\documentstyle[12pt,aasms4,psfig,epsf]{article}

\lefthead{K. Krisciunas {\em et al.}}
\righthead{Five Type Ia Supernovae}

\begin{document}

\title{Optical and Infrared Photometry of the Type Ia
Supernovae 1999da, 1999dk, 1999gp, 2000bk, and 2000ce}
\author{Kevin Krisciunas$^1$,
Mark M. Phillips$^2$,
Christopher Stubbs$^3$,
Armin Rest$^3$,
Gajus Miknaitis$^3$,
Adam G. Riess$^4$,
Nicholas B. Suntzeff$^1$,
Miguel Roth$^2$,
S. E. Persson$^5$, and
Wendy L. Freedman$^5$}
\affil{$^1$Cerro Tololo Inter-American Observatory, Casilla 603,
La Serena, Chile \\
$^2$Las Campanas Observatory, Carnegie Observatories, Casilla 601, La
Serena, Chile \\
$^3$Department of Astronomy, University of Washington, Box 351580,
Seattle, WA 98195$-$1580 \\
$^4$Space Telescope Science Institute, 3700 San Martin Drive,
Baltimore, MD 21218 \\
$^5$Observatories of the Carnegie Institution of Washington, 813
Santa Barbara Street, Pasadena, CA 91101 }
\begin {center}
Electronic mail: kkrisciunas, nsuntzeff@noao.edu \\
mmp, miguel@lco.cl \\
stubbs, rest, gm@astro.washington.edu \\
ariess@stsci.edu \\
persson, wendy@ociw.edu \\
\end {center}

\begin{abstract} 

We present BVRI photometry of the Type Ia supernovae 1999da,
1999dk, 1999gp, 2000bk, and 2000ce, plus infrared photometry
of three of these.  These objects exhibit the full range of
decline rates of Type Ia supernovae.
Combined optical and infrared data show that families
of V $-$ infrared color curves can be used to derive the
host extinction (A$_V$) of these objects.  Existing data
do not yet allow us to construct these loci for all color indices
and supernova decline rates, but the 
V$-$K color evolution is sufficiently uniform that it
allows the determination of host extinction  
over a wide range of supernova decline rates to an accuracy of roughly 
$\pm$ 0.1 mag.  We introduce a new empirical
parameter, the mean I-band flux 20 to 40 days after maximum
light, and show how it is directly related to the decline rate.

\end{abstract}


\section{Introduction}

Over the past decade Type Ia supernovae (SNe) have been shown to be
excellent standardizable candles (Leibundgut 2000, and references therein).
The most popular models rely on the detonation (or deflagration) of a C-O
white dwarf at the Chandrasekhar mass.  Because the amount of fuel in the
explosion is approximately
the same, the resulting explosion has about the same amount of energy.
The optical light curve is powered by the amount of $^{56}$Ni that is
produced, typically 0.6 M$_{\odot}$ (Nomoto, Iwamoto, \& Kishimoto 1997).

However, it is known that Type Ia SNe {\em do} exhibit a range of peak
luminosities, and models bear out that the least luminous examples, such
as SN 1991bg, produce only $\approx$~0.1 M$_{\odot}$ of $^{56}$Ni, while the more
luminous examples, such as SN 1991T, produce $\approx$~1.1 M$_{\odot}$
(Cappellaro et al. 1997; Contardo, Leibundgut, \& Vacca 2000). 
Details of modeling the physics of the explosion, along with a comparison
with the well observed SNe 1991T and 1992A can be found in Pinto \&
Eastman (2000a, 2000b); see also H\"{o}flich, Khokhlov, \& Wheeler (1995).

It is now well known that the luminosities of Type Ia SNe obey a decline rate
relation (Phillips 1993, Phillips et al. 1999).  Hamuy et al. (1996a, 1996b) and
Phillips et al. (1999) have developed a three filter (BVI) template-fitting
procedure, commonly referred to as the ``$\Delta$m$_{15}$(B) method'', for deriving
distances to Type Ia SNe and for estimating the host galaxy dust reddening.  Here
$\Delta$m$_{15}$(B) is the number of B-band magnitudes the SN dims in the first 15
days after the time of maximum light. Those with slower decline rates are
intrinsically brighter at maximum light.  It is important to note that slow
decliners are slow risers; also,  fast risers are fast 
decliners.\footnote[1]{The correlation of rise rate with decline rate is
well known to supernova observers, but one rarely finds the idea expressed
this simply.  The template {\em stretch method} of Perlmutter et al. (1997) uses
templates running from {\em negative} 10 days to +80 days with
respect to
the time of B-band maximum.  By definition, the B-band light is on the
rise for $t <$ 0, since t = 0 is the time of B$_{max}$.
The method of Riess, Press, \& Kirshner (1996), which generates templates
from $t = -$10 to +90 days with respect T(B$_{max}$), 
and that of Phillips et al. (1999) speak of the {\em width} of the light curves
being larger for the slowly declining Type Ia SNe.  All three methods
imply that slow risers are slow decliners, and fast risers are
fast decliners.  Recently, this has been shown to be quantitatively
true by Goldhaber et al. (2001).}

Riess, Press, \& Kirshner (1996, hereafter RPK) and Riess et al. (1998a)
developed the multi-color light curve shape (MLCS) method of fitting the
BVRI photometry of Type Ia SNe.  The key parameter, called $\Delta$, is the
number of V-band magnitudes (at maximum) that a Type Ia SN is brighter 
than ($\Delta < 0$) or fainter than ($\Delta > 0$) some fiducial object.  From the
calibration of the distances of the host galaxies using different methods, a Type Ia SN
with $\Delta$ = 0.00 has a corresponding absolute magnitude at maximum of
M$_V$ = $-$19.46.  The range of $\Delta$ (see RPK) is roughly $-$0.56
(for SN 1991T) to +1.44 mag (for SN 1991bg).

It is worth emphasizing that both the $\Delta$m$_{15}$(B) method
and MLCS use photometry from multiple filters to derive a single
parameter keyed to a specific filter.  The main difference is
that MLCS uses a linear interpolation method to derive a
{\em continuum} of templates, while the $\Delta$m$_{15}$(B) method
uses a {\em finite number} of templates derived from fits to the data
of specific, individual SNe.

Recently, Krisciunas et al. (2000, hereafter Paper I) showed that Type Ia
SNe which are spectroscopically ``normal'' and which have $-0.38 \lesssim
\Delta \lesssim +0.23$ mag exhibit uniform V $-$ near infrared color
curves. We found that V$-$K colors get linearly bluer from 9 days before
B-band maximum until 6 days after T(B$_{max}$), after which they redden
linearly, until roughly 27 days after T(B$_{max}$).
V$-$H colors obey quite similar color evolution.  However, V$-$J colors
exhibit considerably more scatter for objects with comparable reddening;
given the complexity of the spectrum of Type Ia SNe in the J-band (see
Wheeler et al. 1998), it is not surprising that any color index involving
J would behave differently from object to object.

Paper I showed that there exist fiducial color loci which can be used
to determine the total absorption (A$_V$) towards Type Ia SNe, even if
the reddening in the host galaxy is significantly different than
the reddening by dust in our Galaxy.  For example, SN 1999cl in the Virgo Cluster
galaxy M 88 was reddened, with R$_V$ $\equiv$ A$_V$/E(B$-$V)
$\approx$ 1.8, vastly different than the canonical value of 3.1
(Sneden et al. 1978; Rieke \& Lebofsky 1985).  

Because less than 100 Type Ia SNe have well characterized BVRI light
curves, and only a small percentage were observed in the near
infrared (see Meikle 2000 for a summary of the infrared observations),
we have been carrying out a program of such observations of
newly discovered objects. In this paper we provide optical light
curves of SNe 1999da, 1999dk, 1999gp, 2000bk, and 2000ce.  The last
three were also measured in the near infrared.  In addition to
providing finding charts, field star sequences, and supernova
photometry, we also introduce a new parameter which is well correlated
with the intrinsic brightness of these objects.

\section{Observations}

SN 1999da was discovered by Johnson \& Li (1999) from images of 2.4 and
5.4 July 1999 UT.  It occurred in NGC 6411, which Skiff (private 
communication) indicates 
is best classified as a mid-stage lenticular galaxy of type
SA0$^{-}$.  Spectra by Filippenko (1999) and Jha et al. (1999a)
indicated that this would be a subluminous Type Ia SN similar to SNe 1991bg
(Leibundgut et al. 1993) and 1998de (Modjaz et al. 2001).  

SN 1999dk was discovered by Modjaz \& Li (1999) from images of 12.5 and
13.5 August 1999 UT.  It occurred in the Sc galaxy UGC 1087.  

SN 1999gp was discovered in the Sb galaxy UGC 1993 by Papenkova \& Li
(1999) on 23.2 December 1999 UT.  A spectrum taken on 2.2 January 2000
UT and reported by Jha et al. (1999b) indicated, from the absence of a
strong Si II 580 nm feature, that this SN would prove to be overluminous.  
Nevertheless, in this same spectrum, which was obtained $\sim$5 days before
B-band maximum, reasonably strong Si~II 635.5 nm absorption is clearly
present, so this SN would not appear to be as extreme as a 1991T or 1999aa-like
event.

SN 2000bk was discovered in the SA0$^-$ galaxy NGC 4520 by
Armstrong (2000) on 12.0 April 2000.  
 
SN 2000ce was discovered in the SBb galaxy UGC 4195 by Puckett
(2000) on 8.1 May UT.  

On the basis of observations of Landolt (1992) standards on photometric
nights, we determined the V magnitudes and optical colors of field
stars near each supernova.  In the course of our calibrations we
discovered that Landolt's published value of the the V$-$I color of the
star PG0231+051 has a systematic error.  The correct value is
$-$0.350 $\pm$ 0.017.  Stetson (private communication) confirms
this; his value is $-$0.346 $\pm$ 0.018.  Without taking this into
account, our derived V$-$I colors for SN 1999gp would be in error
by up to 0.1 mag.

All of our photometry of SN 1999gp and 2000ce was obtained with the Apache
Point Observatory (APO) 3.5-m telescope. All of the SN 1999dk data were
taken with the University of Washington's Manastash Ridge Observatory
(MRO) 0.76-m telescope.  SN 1999da was observed with both APO and MRO.  
SN 2000bk was observed at APO (BVRIJH photometry) and with the Las Campanas
Observatory 1-m and 2.5-m telescopes (J and H).

In Figs. 1 through 5 we give finding charts for each SN.  Tables
1 through 5 give coordinates of each SN, coordinates of the field
stars, and optical photometry of these field stars.  The field
star photometry is based on averages of four to six photometric
nights of calibration per field.  Differential photometry amongst
the field stars for each supernova indicates no obvious variables.
We find these stars to be constant at the
$\pm$ 0.03 mag level or better, so differential photometry of
the SNe with respect to these stars gives us confidence that
any variability is attributable to the SNe themselves.

Ideally, a supernova occurs at a large enough angular distance from the
luminous regions of its host galaxy and also from the field stars in
our Galaxy, such that simple aperture photometry is accurate.  If the
supernova is superimposed on the light of the host galaxy, aperture
photometry is still sufficiently accurate if the underlying light of
the galaxy is reasonably uniformly distributed.  This can be judged
by first considering the instrumental magnitudes as a function of
aperture size for the isolated field stars and comparing them to the
instrumental magnitudes as a function of aperture for the supernova.  If
both sets of numbers converge at the same rate, then we can likely
trust the aperture photometry of the SN. This is certainly the case
for SNe 1999da and 2000bk, which occurred in the low-luminosity outer
reaches of two early type galaxies.  Aperture photometry also is
justifiable in the cases of SNe 1999dk and 2000ce.  However, it was
clear that SN 1999gp would require reduction via image subtraction,
using templates obtained after the SN had faded sufficiently.  For
image subtraction we used the algorithms and software of Alard
\& Lupton (1998), with additional scripts and improvements written by
Andrew Becker and by us.

Our IR templates for SN 1999gp were obtained on 17 October 2000 and our
BVRI templates were obtained on 27 November 2000, some 284 and 325 days,
respectively, after the date of B-band maximum.  We compared results from
aperture photometry and image subtraction techniques.\footnote[2]{The
B-band photometry required corrections of 0.01 to 0.12 mag, while the
I-band photometry required corrections of 0.03 to 0.43 mag.}  The non-uniform
nature of the underlying galaxy light is clearly a problem with this
object.  Image subtraction was required at all wavelengths.

Our local infrared standards for SN 2000bk were calibrated on seven
photometric nights using the standard star system of Persson et al.
(1998).\footnote[3]{The Las Campanas Observatory data were obtained
with a standard H-band filter, but their J$_s$ (``J-short'') filter
has a narrower bandwidth and a different central wavelength than
the more standard J filter used at Apache Point Observatory.  
We do not have the necessary observations of {\em stars} to give
transformations between these infrared systems, and even if we did, the question 
remains as to the proper transformations required for the reduction of the 
{\em supernova} data.}
The local standards for SN 2000ce were calibrated on five
photometric nights using the list of Hunt et al. (1998).  Local
infrared standards for SN 1999gp were also calibrated with observations
of standards of Hunt et al. (1998); the near infrared magnitudes 
of these stars were found to agree within the errors with preliminary 
values from the 2MASS survey (see Table 11).

For the infrared photometry of SNe 2000bk and 2000ce we compared the results
of aperture photometry and PSF (point spread function) magnitudes obtained
with {\sc daophot} (Stetson 1987, 1990).  This also involved the use of
additional numerical tools developed by one of us (NBS).  The PSF magnitudes
typically gave smoother light curves and color curves, so we have 
adopted the PSF magnitudes for the infrared photometry of these two
objects.  

In Tables 6 through 10 we give BVRI photometry of our five SNe, based on our
calibration of the local secondary ``standards''.  The BVRI photometry is
weighted by the photon statistics of the raw data and the uncertainties
resulting from the calibration of the local secondary standards.  In our
experience, formal uncertainties smaller than $\pm$ 0.010 mag are not to be
taken too literally.

Light curves are shown in Figs. 6 through 10. On the left hand sides
of Figs. 7 through 10 we show the MLCS fits to the BVRI data.  On
the right hand sides we show the BVI templates derived via the
$\Delta$m$_{15}$(B) method.  In the case of SN 2000ce the latter
method finds that two templates (1991T and 1992bc)
fit the data equally well.  This
leads to greater uncertainty than usual regarding the time of B-band
maximum, and the magnitudes at maximum.  This is an inherent problem
if observations of a supernova are not made closer to the time
of maximum light.

  We remind the reader that the MLCS method recognizes that there is
inherent scatter of the V-band magnitudes and B$-$V, V$-$R, and V$-$I
colors of supernovae {\em even if the photometry can be characterized
by the same value of} $\Delta$.  Note the width of the ``grey
snakes'' in Fig. 3 of RPK. MLCS solutions preferentially give greater
weight to the photometry obtained at maximum light.  Deviations of
the templates from the photometry of an {\em individual} supernova is
more a recognition of the inherent scatter amongst objects than a
failure of the MLCS method.

Table 11 gives infrared magnitudes of local field stars.  Table 12
gives near infrared photometry of SNe 1999gp, 2000bk and 2000ce.

We note in passing that we applied color terms, derived from observations
of Landolt (1992) standards, to the BVRI data for the SNe, but no color
terms were applied to the infrared photometry.  Most of the infrared
observations of SN 2000bk were obtained with the Las Campanas 1-m
telescope.  It was on this very telescope that the infrared standard
system of Persson et al. (1998) was defined, so the color terms
would be non-zero only if the optical coatings had greatly changed
over time.

\section{Discussion} 

\subsection{General comments}

In Table 13 we summarize various observational parameters for these
five SNe based on MLCS.  For comparison, in Table 14 we give light curve
parameters (time of B-band maximum, decline rate, best fit magnitudes
at maximum light, host reddening, and distance modulus) based on 
analysis similar to that of Phillips et al. (1999).

The second version of MLCS (Riess et al. 1998a) was limited to $-0.5
\leq \Delta \leq +0.5$.  SN 1999da is a faster decliner than a Type
Ia SN characterized by $\Delta$ = +0.50, so it cannot be fit with
the latest MLCS vectors.  The decline rate $\Delta$m$_{15}$(B) = 1.94
$\pm$ 0.10 is amongst the most rapid known.
Very few fast declining Type Ia SNe like
1991bg (Leibundgut et al. 1993) and 1998de (Modjaz et al. 2001) have
well sampled light curves.  Given the rarity of such objects and the
fact that we began observing SN 1999da before maximum light, our
results will eventually be valuable for the calibration of light
curve fitting schemes for the low luminosity range of Type Ia SNe.

SN 1999gp showed an extremely prominent secondary maximum in the I-band,
which is indicative of an overluminous Type Ia SN.  For this object MLCS
gives $\Delta$ = $-$0.45 $\pm$ 0.10 mag.

SN 2000bk was somewhat underluminous, with $\Delta$ = +0.43 $\pm$ 0.15
mag. As one can see in Fig. 9, it had a reasonably weak I-band secondary
maximum, but a very strong one in the J-band.

SN 2000ce was a moderately luminous SN, with $\Delta$ = $-0.26 \pm
0.17$ mag.  

The earliest spectra obtained of SNe 1999aa, 1999da, and 1999gp predicted
that the first and last were overluminous Type Ia SNe, while SN 1999da was
underluminous.  Subsequent photometry proved that their light curves were
very much as predicted.  This underscores the insights of Nugent et al.
(1995) and Riess et al. (1998b), who suggested  that a spectrum taken within 
a week of maximum light and photometry
obtained on a small number of nights near maximum make it possible to
estimate the intrinsic brightness of a Type Ia SN.  The results of
Paper I also indicate, if such an object is
spectroscopically ``normal'', that H-band or K-band photometry earlier
than a month after maximum light, combined with a V-band light curve, can
give us the extinction (A$_V$) to $\pm$0.1 mag or better.  Future
supernova observing campaigns should certainly take this into account.

In Fig. 11 we show the reddening-corrected absolute magnitudes in BVI
of SNe 1999da, 1999dk, 1999gp, 2000bk, and 2000ce plotted versus 
$\Delta$m$_{15}$(B).  The
fainter crosses correspond to the 41 SNe in Phillips et al. (1999) with
0.01 $< z < $ 0.1.  To these has been added the fast-declining SN 1998de
(Modjaz et al. 2001).  The distance to each SN was calculated from the
redshift of the host galaxy (in the cosmic microwave background frame)
and an assumed value of the Hubble constant of H$_0$ = 65 km s$^{-1}$ Mpc$^{-1}$.
Following Hamuy et al. (1996a), a peculiar velocity term of 600 km s$^{-1}$
has been included in the error bars of the absolute magnitudes.  The  
host galaxy reddening for each SN was calculated using the methods
given in Phillips et al. (1999).  Note that the five SNe from the
present paper are in excellent agreement with the larger sample.

We note that the distance moduli for SNe 1999dk, 1999gp, 2000bk, and 2000ce are, on
average, 0.22 mag larger using MLCS compared to the results based on analysis in the
style of Phillips et al. (1999). For these four objects, this means that
MLCS-derived distances are 11 percent larger and would lead to correspondingly
smaller estimates of the Hubble constant.

\subsection {Extinction and reddening}

The simplest way to test the
``uniformity'' of our V $-$ near infrared color relations given in
Paper I is to plot the V $-$ IR colors of other SNe and derive color 
excesses from a simple upward translation of the  unreddened loci.
If the color excesses from V$-$J, V$-$H, and V$-$K imply the
same value of A$_V$, then all is well.\footnote[4]{As in Paper I,
we adopt the reddening  calibration of Rieke \& Lebofsky (1985),
which gives A$_V$ = 1.393 E(V$-$J) = 1.212 E(V$-$H) = 1.126
E(V$-$K).}  If there are residuals which
correlate with a luminosity-related parameter (i.e. MLCS $\Delta$ or
$\Delta$m$_{15}$(B)), that is evidence that there are families
of loci analogous to BVRI templates for optical photometry.

In Figs. 12, 13, and 14 we plot V $-$ IR colors for SNe 1999gp,
2000bk, and 2000ce.  We also plot our unreddened loci from Paper I, along
with those loci offset by various amounts.

While the V$-$K colors of SN 1999gp are too uncertain to say anything
of substance, the V$-$J and V$-$H colors confirm what we showed in Fig. 12 of
Paper I, that in the lower right region  of the diagrams
the V $-$ near IR colors of unreddened, overluminous, slowly 
declining Type Ia SNe are to be found below the unreddened loci determined
by mid-range decliners.  

Given the generally low dust content of early-type galaxies and the
location of SN 2000bk in the outskirts of its host, one might expect
absorption by dust in its host to be minimal. MLCS gives A$_V$ = 0.28
$\pm$ 0.20 along the line of sight towards this SN.  The Schlegel et al.
(1998) reddening maps of our Galaxy indicate that along the line of sight
to SN 2000bk the Galactic reddening is E(B$-$V) = 0.025 mag, so 0.08 mag
of A$_V$ is due to dust in our Galaxy. Analysis similar to that of
Phillips et al. (1999) indicates a host reddening of E(B$-$V) = 0.10 $\pm$
0.04.  Assuming R$_V$ = 3.1 for dust in our Galaxy and the host of SN
2000bk, A$_V$ = 0.39 $\pm$ 0.12 mag. Thus, the extinction suffered by this
SN would appear to be small, but not zero.

The {\em shape} of the V$-$J color curve of SN 2000bk clearly does
not correspond to a simple upward translation of the unreddened locus.  Even
if we restrict the fit to data earlier than 18 days after B-band maximum, the
implied color excess is E(V$-$J) = 0.62 $\pm$ 0.03 mag.  Using a standard
dust model (Eqs. 5, 6, and 7 of Paper I) implies A$_V$
= 0.87 $\pm$ 0.08 mag, clearly much larger than the values of 0.28
$\pm$ 0.20 and 0.39 $\pm$ 0.12 quoted above. The V$-$H colors of SN 2000bk
imply A$_V$ = 0.54 $\pm$ 0.04 mag, which is also larger than the values
based on BVRI data.  However, the {\em slope} of the V$-$H color curve 
(0.0851 $\pm$ 0.0059 mag d$^{-1}$) is, within the errors, identical
to that of our unreddened locus (0.0840 $\pm$ 0.0042).

If we assume that the color evolution of SN 2000ce is similar to the SNe
studied in Paper I, the V$-$J, V$-$H and V$-$K data give estimates of
A$_V$ of 1.80 $\pm$ 0.44, 1.33 $\pm$ 0.13, and 1.65 $\pm$ 0.12 mag,
respectively, where we have included the contribution of the uncertainty
of the time of B-band maximum to the error bars.  {\em However}, the few
V$-$J points only fit the template if we shift the data 3.0 days to the
left in Fig. 12 (i.e. adopt T(B$_{max})$ 3 days later than the value
given by MLCS).  There is evidence (see below and Fig. 15) that we should
use a bluer V$-$H template for this object than that of Paper I. A$_V$
derived from V$-$K photometry is in excellent agreement with the value of
A$_V$ = 1.67 $\pm$ 0.20 given by MLCS.  This is to say that the
photometry of SN 2000ce points to: 1) an ongoing suspicion that V$-$J is
not a ``well behaved'' color index; 2) our V$-$H template from Paper I is
not applicable to as wide a range of $\Delta$ as we previously thought;  
3) the analysis confirms that A$_V$ for a reddened Type Ia SN can be
determine from V$-$K colors with an uncertainty of roughly $\pm$ 0.1 mag.  
In the case of SN 2000ce the uncertainty of T(B$_{max}$) contributes
as much to the error budget as the photometric errors or the uncertainty
of the ratio of total to selective absorption [R$_{\lambda_{1}}$ =
A$_{\lambda_{1}}$/E($\lambda_1 - \lambda_2$)].

Since one of our goals is to be able to use optical and infrared
data to determine values of A$_V$ (and hence the distances) of Type
Ia SNe over the full range of decline rates, but a sufficient number
of well sampled IR light curves does not yet exist, we proceed
tentatively as follows. SN 1998bu data shown in Paper I and
preliminary SN 1999ac data (Phillips et al., in preparation) show
that the local minimum in the V$-$H and V$-$K colors a week after
T(B$_{max}$) has a more gradual change of slope than the
two-straight-line loci used in Paper I.  We remind the reader that
these loci are based on 4 supernovae unreddened in their hosts, two
with minimal reddening, and two with significant reddening.  If we
fit fourth order curves to the ``compacted'' data shown in Fig. 10
of Paper I and subtract the derived color excesses E(V$-$J),
E(V$-$H), and E(V$-$K) of SN 1998bu (Table 10 of Paper I) from the
data, we would expect the deviations of the dereddened SN 1998bu
data from these new loci to be very nearly zero, and in fact they
are.  The mean deviations are $-$0.025 $\pm$ 0.031, +0.013 $\pm$
0.020 and +0.007 $\pm$ 0.019 mag, respectively, for V$-$J, V$-$H and
V$-$K.  We can then use the BVRI photometry and assumed values of IR
to V-band extinction to deredden the V$-$J, V$-$H and V$-$K data of
SNe 1999aa, 1999ac, 1999gp, 2000bk, 2000ce.\footnote[5]{We have
restricted the data of SNe 1999aa, 1999ac, 1999gp, 2000bk, and
2000ce to $0 \leq t \leq 27$ days after the time of B-band maximum.}

In Fig. 15 we show the mean deviations from the fourth order
unreddened loci of the dereddened V $-$ near IR colors of six
supernovae.  Graphically, this confirms our principal finding from
Paper I, in particular, that V$-$K colors of Type Ia SNe over a
rather wide range of $\Delta$ can be fitted by the same 
locus, adjusted only by the V$-$K color excess of the object.
SN 2000bk and the peculiar SN 1986G (see Fig. 12 of Paper I)
show that fast decliners are intrinsically redder in V $-$ near IR
colors than mid-range decliners.  SNe 1999aa and 1999gp show that
slow decliners are intrinsically bluer than the mid-range decliners.
The different values of A$_V$ for SN 2000ce derived from V$-$H and
V$-$K data can be explained by the fact that at $\Delta = -0.26$ we
need a bluer unreddened locus for V$-$H (and/or one with a different
slope from 9 to 27 days after T(B$_{max}$)) than the one that is our
present reference.

The data of the present paper are consistent with the prediction of Paper
I, namely that fast decliners, mid-range decliners, and slow decliners
require families of curves to delineate their color evolution.  Still,
the V $-$ near IR colors of mid-range decliners may be characterized by
thin enough ``grey snakes'' (see RPK, Fig. 3) that they can be used as an
excellent basis for determining A$_V$.

\subsection {The secondary hump in the red}

One of the most characteristic features of the light curves of Type
Ia SNe is the secondary maximum in the far red and near infrared
(IJHK bands).  Elias et al. (1981, 1985) first showed this for the
JHK bands. For the I-band it was first discussed by Ford et al.
(1993), and in 1993 at the Xian, China, meeting on supernovae
(Suntzeff 1996). Subsequently, Hamuy et al. (1996d) provided evidence
that the I-band secondary maximum is stronger and occurs later for
the slowly declining Type Ia SNe. In a parallel study, RPK also
showed graphically that the secondary I-band maximum is very strong
for the most luminous Type Ia SNe, while there is no secondary
maximum for the least luminous ones. H\"{o}flich, Khokhlov, \&
Wheeler (1995) explain this as a temperature/radius effect.  To a
first approximation, the infrared luminosity is proportional to the
area of the expanding fireball, the effective temperature, and a
dilution factor for scattering dominated atmospheres.  If the
photospheric radius (R$_{ph}$) is still increasing after maximum, a
secondary maximum occurs due to the R$^{2}_{ph}$ term. For the
subluminous examples the secondary maximum is not seen because it
merges with the primary maximum.  For additional discussion and
references see \S3.1 of Riess et al. (2000) and \S4.1 of Pinto
\& Eastman (2000b).

We pose the question: to what extent is the luminosity at maximum of a
Type Ia SN correlated with the strength of the secondary I-band maximum?  
For this we need well sampled I-band light curves. In Fig. 16 we show the
I-band photometry of SN 1999aa from Paper I, with the I magnitudes
converted to flux with respect to the I-band maximum.  We have fitted a
higher order polynomial to the data and determined the mean flux (based
on an {\em integration} of the polynomial) from 20 to 40 days after the
time of B-band maximum.  In this example $\langle$ I $\rangle _{20-40}$ =
0.617.  For a well sampled light curve and typical photometric
uncertainties, this parameter should be accurate to
$\pm$0.03 or better.  In fact, two independent sets of data for
SN 1996X (Riess et al. 1999; R. Covarrubias et al., unpublished) give
values of $\langle$ I $\rangle _{20-40}$ = 0.516 and 0.471, respectively,
indicating an uncertainty of $\pm$ 0.023 for the mean. A third data
set for SN 1996X (Salvo et al. 2001) gives $\langle$ I $\rangle _{20-40}$ =
0.521, but their I-band photometry is particularly ragged at the time
of the secondary I-band maximum.

In Fig. 17 we show $\langle$ I $\rangle _{20-40}$ vs.
$\Delta$m$_{15}$(B) for 23 SNe.  We include four SNe from the RPK
training set: 1990N and 1991T (Lira et al. 1998), 1991bg (Leibundgut
et al. 1993), plus 1992A (Suntzeff et al., unpublished). SN 1992K
(Hamuy et
al. 1994) was a subluminous, rapid decliner. We include five
other SNe from
the Cal\'{a}n/Tololo survey (Hamuy et al. 1996c): 1992al, 1992bc,
1992bo, 1993H, and 1993O; seven of the 22 SNe from Riess et al.
(1999): 1994M, 1994ae, 1995D, 1995E, 1995ac, 1995al, and 1996X; SN
1998bu (Jha et al. 1999c); SN 1999aa (Paper I); plus four of the five
SNe presented in this paper.  SN 2000ce was not observed early enough
to get an accurate estimate of the I-band maximum.  These 23
SNe have a median total color excess (i.e. sum of Galactic 
and host galaxy contributions) of E(B$-$V) = 0.104 mag, with only
SNe 1995E and 1998bu having values greater than 0.2.

Except for SNe 1992bc and 1994M, the points plotted in Fig. 17
form a tight relation. In the top panel of Fig. 18 we show the
BVI light curves of SNe 1994ae and 1992bc.  These objects have
essentially identical decline rates in B and V, but SN 1992bc has
a considerably weaker I-band secondary maximum.  In the bottom
panel of Fig. 18 we similarly show SNe 1992A and 1994M, where the
latter shows a much stronger I-band maximum.

Since the SN 1992bc data were obtained with CTIO telescopes and
detectors and reduced in the same manner as the data for the 10
other SNe plotted in Fig. 17 observed at CTIO, there is no reason
to suspect I-band calibration problems to be the cause of its
anomalous I-band secondary maximum.  Similarly, SN 1994M was
observed with the same equipment and reduced in the same manner
as the 7 other SNe with Center for Astrophysics (CfA)  data
plotted in Fig. 17. The excellent agreement of the $\langle$ I
$\rangle _{20-40}$ values calculated from the CfA and CTIO data
sets for SN 1996X further demonstrates that these outliers are
real.  Pinto \& Eastman (2000b, \S 4.1) suggest that the details
of the secondary maximum tell us something about the ignition
conditions of the explosion, in particular the neutron-rich
isotopes that make up the non-radioactive core.

The solid curve in Fig. 17 is a third order polynomial fit.  If we
eliminate SNe 1992bc and 1994M from the fit, the {\sc rms} residual is
only $\pm$ 0.093 in $\Delta$m$_{15}$(B). Figs. 17 and 18 show that the
luminosity at maximum is directly related to the strength of the I-band
secondary maximum, but that there are clearly exceptions to the rule.  

The $\langle$ I $\rangle _{20-40}$ relationship
and the demonstrated usefulness of the V$-$H and V$-$K color curves
for determining A$_V$ should provide ample motivation for the gathering
of optical and infrared data for Type Ia SNe.

\section{Conclusions}

We have presented optical photometry of SNe 1999da, 1999dk, 1999gp, 
2000bk, and 2000ce.  We obtained reasonably well-sampled
J-band and H-band light curves of SN 2000bk, along with a small amount 
of infrared photometry of SNe 1999gp and 2000ce.

A combination of V-band and infrared photometry allows one to determine the 
extinction suffered by Type Ia SNe in their hosts.  At present, however,
the data only exist to do this for Type Ia SNe in the mid-range of
decline rates.   

By introducing a new empirical parameter, $\langle$ I $\rangle _{20-40}$,
we provide a useful way of parameterizing the strength of the I-band
secondary maximum of Type Ia SNe.  This is directly related to the 
decline rate, and hence, to the intrinsic brightness of these objects.  

\vspace {1 cm}

\acknowledgments

This paper is based in part on observations obtained with the Apache
Point Observatory 3.5-meter telescope, which is owned and operated
by the Astrophysical Research Consortium.  {\sc iraf} is a product
of the National Optical Astronomy Observatories, which is operated
by the Association of Universities for Research in Astronomy, Inc.,
under cooperative agreement with the National Science Foundation.

We thank Eugene Magnier and Alan Diercks for the use of their infrared
data reduction software. Andrew Becker provided scripts for an image
subtraction pipeline.  Brian Skiff kindly determined the coordinates
of the field stars in Table 1 through 5.  We thank Russet
McMillan and Camron Hastings for observing support. We also made use
of Simbad, a database of the Centre de Donn\'{e}es
astronomiques de Strasbourg.

This work was supported by NSF grant AST-9512594.  C. Stubbs also
acknowledges the generous support of the McDonnell Foundation.


\clearpage
\begin{deluxetable}{ccccccc}
\tablewidth{0pc}
\tablecaption{NGC 6411 Photometric Sequence for SN 1999da}
\tablehead{
\colhead{$\star$} & \colhead{$\alpha$ (2000)$^a$} &
\colhead{$\delta$ (2000)$^a$} & \colhead {V} & \colhead {B$-$V} &
\colhead{V$-$R} & \colhead{V$-$I} }
\startdata
 SN & 17:35:23.0 & +60:48:49 &                &                &               &                \nl
  2 & 17:35:25.9 & +60:47:37 & 18.311 (0.014) &  0.724 (0.022) & 0.447 (0.020) &  0.842 (0.023) \nl
  3 & 17:35:29.8 & +60:47:24 & 17.861 (0.015) &  0.808 (0.016) & 0.463 (0.011) &  0.852 (0.009) \nl
  4 & 17:35:48.9 & +60:48:29 & 18.171 (0.040) &  1.368 (0.120) & 1.054 (0.052) &  2.360 (0.076) \nl
  5 & 17:35:33.5 & +60:50:25 & 16.904 (0.018) &  1.392 (0.015) & 0.824 (0.018) &  1.590 (0.022) \nl
  6 & 17:35:23.8 & +60:52:42 & 14.656 (0.002) &  0.971 (0.006) & 0.547 (0.003) &  1.075 (0.003) \nl
  7 & 17:35:20.1 & +60:50:09 & 15.121 (0.007) &  0.773 (0.004) & 0.454 (0.008) &  0.838 (0.010) \nl
\nl
\enddata
\tablenotetext{a} {The field star coordinates were derived from the USNO-A2.0 catalog.
Photometry of stars 2, 3, 5, and 7 were calibrated with the Apache Point
Observatory 3.5-m telescope on one photometric night. Stars 4 and 6 were calibrated with
observations from the Manastash Ridge Observatory 0.76-m telescope.}
\end{deluxetable}

\begin{deluxetable}{ccccccc}
\tablewidth{0pc}
\tablecaption{UGC 1087 Photometric Sequence for SN 1999dk}
\tablehead{
\colhead{$\star$} & \colhead{$\alpha$ (2000)$^a$} &
\colhead{$\delta$ (2000)$^a$} & \colhead {V} & \colhead {B$-$V} &
\colhead{V$-$R} & \colhead{V$-$I} }
\startdata
 SN &  1:31:26.9 & +14:17:06 &                &                 &                &                \nl
  2 &  1:31:35.1 & +14:14:00 & 13.555 (0.009) &   0.705 (0.016) &  0.389 (0.020) &  0.747 (0.021) \nl
  3 &  1:31:43.5 & +14:15:31 & 14.374 (0.011) &   0.757 (0.012) &  0.416 (0.020) &  0.785 (0.022) \nl
  4 &  1:31:24.0 & +14:18:31 & 17.207 (0.008) &   0.628 (0.017) &  0.332 (0.032) &  0.665 (0.046) \nl
  5 &  1:31:24.5 & +14:19:18 & 17.294 (0.017) &   1.374 (0.054) &  0.861 (0.034) &  1.625 (0.037) \nl
 \nl 
\enddata 
\tablenotetext{a} {Field star coordinates are from the
USNO-A2.0 catalog.}
\end{deluxetable}

\begin{deluxetable}{ccccccc}
\tablewidth{0pc}
\tablecaption{UGC 1993 Photometric Sequence for SN 1999gp}
\tablehead{
\colhead{$\star$} & \colhead{$\alpha$ (2000)$^a$} &
\colhead{$\delta$ (2000)$^a$} & \colhead {V} & \colhead {B$-$V} &
\colhead{V$-$R} & \colhead{V$-$I} }
\startdata

SN &  2:31:39.2 & +39:22:52 &                &                &                &                \nl
 2 &  2:31:37.5 & +39:23:37 & 16.823 (0.006) &  0.952 (0.009) &  0.489 (0.010) &  1.042 (0.013) \nl
 3 &  2:31:45.6 & +39:22:46 & 17.410 (0.004) &  0.655 (0.007) &  0.328 (0.009) &  0.778 (0.004) \nl
 4 &  2:31:40.6 & +39:21:47 & 17.986 (0.006) &  0.693 (0.005) &  0.364 (0.007) &  0.807 (0.012) \nl
 5 &  2:31:40.0 & +39:21:14 & 17.265 (0.006) &  0.595 (0.016) &  0.332 (0.008) &  0.776 (0.009) \nl

 6 &  2:31:37.4 & +39:21:54 & 16.862 (0.002) &  0.882 (0.007) &  0.470 (0.010) &  1.018 (0.014) \nl
 7 &  2:31:34.1 & +39:21:31 & 18.309 (0.009) &  1.242 (0.005) &  0.711 (0.011) &  1.456 (0.010) \nl
 8 &  2:31:33.0 & +39:23:07 & 18.959 (0.011) &  1.076 (0.016) &  0.618 (0.011) &  1.315 (0.017) \nl
 9 &  2:31:31.0 & +39:24:06 & 15.821 (0.010) &  0.569 (0.006) &  0.313 (0.012) &  0.758 (0.011) \nl
10 &  2:31:39.0 & +39:21:46 & 14.499 (0.002) &  0.576 (0.008) &  0.286 (0.007) &  0.655 (0.029) \nl
\enddata
\tablenotetext{a} {Field star coordinates were determined from the USNO-A2.0 catalog.}
\end{deluxetable}

\begin{deluxetable}{ccccccc}
\tablewidth{0pc}
\tablecaption{NGC 4520 Photometric Sequence for SN 2000bk}
\tablehead{
\colhead{$\star$} & \colhead{$\alpha$ (2000)$^a$} &
\colhead{$\delta$ (2000)$^a$} & \colhead {V} & \colhead {B$-$V} &
\colhead{V$-$R} & \colhead{V$-$I} }
\startdata
SN &  12:33:53.9 & $-$07:22:43 &             &                &                &                \nl
 1 &  12:33:55.3 & $-$07:22:44 & 17.882 (0.006) &  0.982 (0.013) & 0.580 (0.002) & 1.086 (0.009) \nl
 2 &  12:33:48.7 & $-$07:23:37 & 16.462 (0.002) &  0.510 (0.009) & 0.348 (0.004) & 0.690 (0.014) \nl
 3 &  12:33:53.7 & $-$07:23:57 & 18.282 (0.004) &  0.680 (0.012) & 0.429 (0.007) & 0.860 (0.010) \nl
 4 &  12:33:59.2 & $-$07:23:11 & 16.192 (0.004) &  0.456 (0.007) & 0.295 (0.002) & 0.582 (0.011) \nl
 5 &  12:34:00.7 & $-$07:21:55 & 18.340 (0.002) &  0.603 (0.011) & 0.386 (0.004) & 0.748 (0.013) \nl
 6 &  12:33:58.9 & $-$07:21:38 & 18.095 (0.005) &  0.695 (0.014) & 0.442 (0.003) & 0.854 (0.014) \nl
 7 &  12:33:56.7 & $-$07:20:52 & 17.744 (0.003) &  1.530 (0.016) & 0.994 (0.005) & 2.054 (0.007) \nl
 8 &  12:33:48.3 & $-$07:22:35 & 15.000 (0.005) &  0.602 (0.012) & 0.365 (0.005) & 0.719 (0.012) \nl
\enddata
\tablenotetext{a} {Field star coordinates were determined from the USNO-A2.0 catalog.}
\end{deluxetable}

\begin{deluxetable}{ccccccc}
\tablewidth{0pc}
\tablecaption{UGC 4195 Photometric Sequence for SN 2000ce}
\tablehead{
\colhead{$\star$} & \colhead{$\alpha$ (2000)$^a$} &
\colhead{$\delta$ (2000)$^a$} & \colhead {V} & \colhead {B$-$V} &
\colhead{V$-$R} & \colhead{V$-$I} }
\startdata
SN &  8:05:09.5 & +66:47:15 &             &                &                &    \nl
 1 &  8:05:18.8 & +66:47:06 & 15.725 (0.002) &  0.813 (0.005) &   0.448 (0.013) &  0.870 (0.007) \nl
 2 &  8:05:07.6 & +66:46:01 & 15.829 (0.002) &  0.694 (0.004) &   0.413 (0.012) &  0.821 (0.006) \nl
 3 &  8:05:18.4 & +66:45:12 & 16.815 (0.003) &  0.838 (0.022) &   0.461 (0.015) &  0.893 (0.011) \nl
 4 &  8:05:18.8 & +66:46:03 & 17.342 (0.005) &  0.566 (0.006) &   0.338 (0.012) &  0.691 (0.008) \nl
 5 &  8:05:27.0 & +66:46:37 & 16.043 (0.003) &  1.123 (0.005) &   0.641 (0.010) &  1.192 (0.007) \nl
 6 &  8:05:14.6 & +66 47:42 & 18.337 (0.014) &  1.682 (0.089) &   1.220 (0.018) &  2.759 (0.019) \nl

\enddata
\tablenotetext{a} {Field star coordinates were determined from the USNO-A2.0 catalog.}
\end{deluxetable}

\begin{deluxetable}{cccccc}
\tablewidth{0pc}
\tablecaption{BVRI Photometry of SN 1999da}
\tablehead{
\colhead{JD $-$ 2,451,000} & \colhead {Obs$^a$} & \colhead{V} &
\colhead{B$-$V} & \colhead{V$-$R} &
\colhead{V$-$I} }
\startdata
  366.7340 & MRO & 16.543 (0.048) &  0.810 (0.113) & $-$0.003 (0.064) & $-$0.082 (0.068) \nl     
  367.8058 & MRO & 16.552 (0.014) &  0.480 (0.025) &    0.266 (0.017) &  0.130 (0.026) \nl
  368.7848 & MRO & 16.462 (0.008) &  0.516 (0.015) &    0.234 (0.012) &  0.161 (0.020) \nl
  379.7176 & MRO & 16.674 (0.012) &                &    0.431 (0.014) &  0.599 (0.020) \nl
  383.6671 & APO & 17.163 (0.005) &  1.499 (0.006) &    0.565 (0.005) &  0.912 (0.006) \nl
  388.8553 & MRO & 17.423 (0.032) &  1.753 (0.152) &    0.364 (0.042) &  0.729 (0.041) \nl
  392.6838 & APO & 17.827 (0.026) &                &                  &  0.914 (0.028) \nl
  413.6875 & MRO & 18.800 (0.044) &  1.036 (0.132) &    0.313 (0.063) &  0.714 (0.061) \nl 
 \nl
\enddata
\tablenotetext{a} {The MRO data were derived with respect to stars 3, 5, 6 and 7 of
the NGC 6411 photometric sequence, while the APO data were derived with respect
to stars 3, 5 and 7.}
\end{deluxetable}

\clearpage
\begin{table*}
\caption{BVRI Photometry of SN 1999dk$^a$}
\vspace{2 mm}
\begin{tabular}{ccccc}
\hline
\hline
JD $-$ 2,451,000 & V & B$-$V & V$-$R & V$-$I \\
\hline
  412.9809 &  14.963 (0.011) &                &  0.006 (0.016) & $-$0.220 (0.021) \\   
  413.8730 &  14.986 (0.008) &  0.082 (0.013) &  0.028 (0.014) & $-$0.229 (0.017) \\
  416.9244 &  14.923 (0.011) &  0.113 (0.019) & $-$0.002 (0.017) & $-$0.368 (0.022) \\   
  425.8880 &  15.216 (0.009) &  0.497 (0.016) & $-$0.085 (0.016) & $-$0.475 (0.025) \\
  426.8637 &  15.270 (0.008) &  0.476 (0.015) & $-$0.122 (0.016) & $-$0.458 (0.025) \\
  437.8778 &  15.820 (0.010) &  1.088 (0.024) &  0.225 (0.018) &  0.256 (0.022) \\
  442.8296 &  16.136 (0.011) &  1.297 (0.026) &  0.342 (0.017) &  0.438 (0.025) \\
  463.8527 &  16.966 (0.026) &  1.100 (0.059) &  0.251 (0.035) &  0.317 (0.046) \\
  464.7227 &  17.053 (0.030) &  0.900 (0.050) &  0.425 (0.041) &  0.309 (0.061) \\
  466.7203 &  16.998 (0.023) &  0.936 (0.045) &  0.345 (0.033) &  0.393 (0.051) \\
  469.8508 &  17.154 (0.020) &  1.016 (0.044) &  0.303 (0.033) &  0.289 (0.041) \\
 \hline
\end{tabular}
\begin{tabular}{l}
$^a$The data were derived with respect to the four stars of the UGC 1087  \\
photometric sequence. 
\end{tabular}
\end{table*}

\begin{table*}
\caption{BVRI Photometry of SN 1999gp$^a$}
\vspace{2 mm}
\begin{tabular}{ccccc}
\hline
\hline
JD $-$ 2,451,000 & V & B$-$V & V$-$R & V$-$I \\
\hline
  542.5664 &  16.626 (0.006) &  0.006 (0.005) &  0.034 (0.005) &  \\
  544.5629 &  16.424 (0.006) &  0.009 (0.004) &  0.029 (0.004) &  0.019 (0.004) \\   
  547.7250 &  16.224 (0.006) &  0.039 (0.004) & $-$0.011 (0.005) & $-$0.102 (0.007) \\   
  552.5816 &  16.140 (0.006) &  0.146 (0.004) &  0.022 (0.003) & $-$0.278 (0.004) \\   
  573.5920 &  17.020 (0.006) &  0.881 (0.004) &  0.122 (0.007) &  0.103 (0.005) \\   
  578.5712 &  17.199 (0.008) &  1.118 (0.005) &  0.266 (0.006) &  0.366 (0.006) \\   
  581.5773 &  17.326 (0.008) &  1.207 (0.004) &  0.358 (0.004) &  0.534 (0.004) \\   
  584.7681 &  17.467 (0.008) &  1.263 (0.007) &  0.377 (0.006) &  0.680 (0.007) \\   
  588.5806 &  17.709 (0.008) &  1.157 (0.014) &  0.453 (0.007) &  0.766 (0.011) \\   
  601.5890 &  18.242 (0.008) &  0.973 (0.011) &  0.321 (0.008) &  0.598 (0.010) \\   
  614.5984 &  18.588 (0.008) &  0.938 (0.009) &  0.283 (0.009) &  0.370 (0.016) \\   
  616.5987 &  18.642 (0.010) &  1.075 (0.041) &                &  0.527 (0.024) \\   
  630.6122 &  18.950 (0.008) &  0.721 (0.012) &  0.177 (0.009) &  0.254 (0.015) \\ 
 \hline
\end{tabular}
\begin{tabular}{l}
$^a$The data were derived with respect to stars 2, 3, 6, 9, and 10 of the  \\
UGC 1993 photometric sequence. 
\end{tabular}
\end{table*}

\begin{table*}
\caption{BVRI Photometry of SN 2000bk$^a$}
\vspace{2 mm}
\begin{tabular}{ccccc}
\hline
\hline
JD $-$ 2,451,000 & V & B$-$V & V$-$R & V$-$I \\
\hline
  658.6877 &  17.289 (0.003) &  0.746 (0.006) &  0.086 (0.004) & $-$0.071 (0.008)  \\
  660.6091 &  17.431 (0.007) &  0.859 (0.015) &  0.123 (0.009) & $-$0.029 (0.014)  \\
  665.8727 &  17.832 (0.005) &  1.252 (0.013) &  0.371 (0.006) &  0.473 (0.012)  \\
  672.6898 &  18.425 (0.003) &  1.273 (0.009) &  0.559 (0.004) &  0.898 (0.007)  \\
  675.6409 &  18.672 (0.005) &  1.206 (0.012) &  0.526 (0.006) &  0.867 (0.008)  \\
  684.6444 &  19.061 (0.006) &  1.229 (0.016) &  0.405 (0.008) &  0.631 (0.011)  \\
  686.6370 &  19.164 (0.006) &  1.192 (0.020) &  0.427 (0.008) &  0.642 (0.013)  \\
  689.6738 &  19.261 (0.005) &  1.126 (0.010) &  0.398 (0.007) &  0.584 (0.011)  \\
  693.6645 &  19.401 (0.005) &  1.113 (0.012) &  0.306 (0.007) &  0.521 (0.014)  \\
 \hline
\end{tabular}
\begin{tabular}{l}
$^a$The data were derived with respect to stars 1, 2, 4, 7, and 8 of the  \\
NGC 4520 photometric sequence. 
\end{tabular}
\end{table*}

\begin{table*}
\caption{BVRI Photometry of SN 2000ce$^a$}
\vspace{2 mm}
\begin{tabular}{ccccc}
\hline
\hline
JD $-$ 2,451,000 & V & B$-$V & V$-$R & V$-$I \\
\hline
  675.6212 &  16.864 (0.005) &  0.778 (0.012) &  0.297 (0.009) &  0.272 (0.007) \\
  684.6168 &  17.442 (0.018) &  1.081 (0.036) &  0.392 (0.022) &  0.638 (0.023) \\
  686.6237 &  17.467 (0.011) &  1.259 (0.043) &  0.338 (0.015) &  0.699 (0.015) \\
  689.6343 &  17.616 (0.004) &  1.489 (0.010) &  0.492 (0.008) &  0.900 (0.007) \\
  693.6401 &  17.796 (0.005) &  1.640 (0.017) &  0.601 (0.009) &  1.117 (0.008) \\
  707.6451 &  18.515 (0.015) &  1.665 (0.065) &  0.682 (0.021) &  1.316 (0.021) \\
  711.6397 &  18.671 (0.017) &  1.825 (0.071) &  0.636 (0.021) &  1.271 (0.020) \\

 \hline
\end{tabular}
\begin{tabular}{l}
$^a$The data were derived with respect to stars 1, 2, 4, and 5 of the UGC 4195 \\
photometric sequence. 
\end{tabular}
\end{table*}

\begin{table*}
\caption{Infrared Magnitudes of Field Stars}
\vspace{2 mm}
\begin{tabular}{cccccc}
\hline
\hline
Field & Star & J & H & K$^{\prime}$ & N$_{obs}$(J,H,K)\\
\hline
SN 1999gp$^a$ &  6  &  15.16(0.02) & 14.68(0.02) & 14.56(0.02) & 2,1,2 \\
          & 10  &  13.34(0.02) & 13.08(0.02) & 13.03(0.02) & 2,1,2 \\
\\
SN 2000bk &  1  &  16.130(0.010) & 15.602(0.012) &         & 7,7,0    \\
          &  2  &  15.357(0.013) & 14.995(0.013) &         & 7,7,0    \\
          &  8  &  13.802(0.007) & 13.448(0.010) &         & 7,7,0    \\
\\
SN 2000ce &  1  &  14.220(0.024) & 13.830(0.020) & 13.800(0.026) & 3,5,4 \\
          &  2  &  14.468(0.030) & 14.069(0.023) & 14.109(0.028) & 2,4,3 \\
          &  6  &  14.136(0.024) & 13.531(0.020) & 13.338(0.025) & 3,5,4 \\
\hline

\end{tabular}
\begin{tabular}{l}
$^a$    Preliminarily 2MASS  values for star 6 are J = 15.13 $\pm$ 0.05, H = 14.62 $\pm$ 0.06, \\
K$_s$ = 14.53 $\pm$ 0.10.  For star 10 2MASS gives J = 13.35 $\pm$ 0.02, H = 13.10 $\pm$ 0.03, \\
and K$_s$ = 13.08 $\pm$ 0.06. (See  http://irsa.ipac.caltech.edu.)
\end{tabular}
\end{table*}

\begin{deluxetable}{ccccccc}
\tablewidth{0pc}
\tablecaption{Infrared Photometry of SNe 1999gp, 2000bk and 2000ce}
\tablehead{   \colhead{SN} &
\colhead{JD$-$ 2,451,000} & \colhead {Obs$^a$} & \colhead{J$^b$} &
\colhead{H} & \colhead{K$^{\prime}$} &
\colhead{K$^c$} }
\startdata
1999gp & 539.57   &  APO & 17.31 (0.06) &               & 17.74 (0.25) &              \nl
     & 561.66   &  APO &              & 17.40 (0.13) & 16.71 (0.11) & 16.51 (0.15) \nl
     & 564.82   &  APO & 18.67 (0.19) &  17.45 (0.14) & 17.27 (0.18) & 17.22 (0.23) \nl
     & 568.79   &  APO & 18.29 (0.12) &  17.16 (0.11) & 17.18 (0.17) & 17.19 (0.22) \nl
\nl
2000bk & 653.77 & LCO1 & 17.64 (0.02) & 17.60 (0.04) & & \nl
     & 654.73 & LCO1 & 17.81 (0.03) &   17.68 (0.04) & & \nl
     & 658.67 & LCO2 & 18.38 (0.03) &   17.74 (0.03) & & \nl
     & 659.63 & LCO2 & 18.41 (0.03) &   17.65 (0.03) & & \nl
     & 660.59 & LCO2 & 18.36 (0.03) &   17.61 (0.03) & & \nl
     & 660.75 & APO  & 18.28 (0.12) &   17.50 (0.07) & & \nl
     & 661.73 & LCO1 & 18.66 (0.04) &   17.73 (0.05) & & \nl
     & 662.68 & LCO1 & 18.45 (0.04) &   17.63 (0.05) & & \nl
     & 663.63 & LCO1 & 18.50 (0.05) &   17.58 (0.05) & & \nl
     & 664.59 & LCO1 & 18.53 (0.04) &                & & \nl
     & 665.65 & LCO1 & 18.48 (0.05) &   17.65 (0.05) & & \nl
     & 665.84 & APO  &              &   17.50 (0.06) &    &   \nl
     & 666.65 & LCO1 & 18.27 (0.04) &   17.58 (0.04) & & \nl
     & 667.60 & LCO1 & 18.22 (0.04) &   17.45 (0.05) & & \nl
     & 672.76 & APO  & 17.85 (0.08) &              &    &   \nl
     & 676.61 & LCO2 & 18.39 (0.02) &   17.94 (0.04) & & \nl
     & 681.65 & LCO2 & 18.92 (0.04) &   18.15 (0.05) & & \nl
     & 683.57 & LCO2 & 19.08 (0.04) &   18.20 (0.05) & & \nl
     & 686.51 & LCO2 & 19.35 (0.08) &                & & \nl
\nl
2000ce & 675.75 &  APO & 16.82 (0.03) & 16.46 (0.03) & 16.16 (0.04) & 16.08 (0.05) \nl
     & 684.72 &  APO & 17.57 (0.04) & 16.42 (0.03) & 16.13 (0.04) & 16.04 (0.05) \nl
     & 686.71 &  APO & 17.44 (0.04) & 16.25 (0.03) & 16.05 (0.04) & 16.00 (0.05) \nl
     & 689.77 &  APO &               & 16.32 (0.03) & 16.11 (0.07) & 16.20 (0.07) \nl
     & 693.74 &  APO &               & 16.12 (0.03) & 16.01 (0.04) & 15.98 (0.05) \nl
\enddata
\tablenotetext{a} {APO = Apache Point Observatory 3.5-m; LCO1 = Las Campanas 1-m;
LCO2 = Las Campanas 2.5-m.
$^b$The LCO observations were made with a J$_s$ filter.
$^c$Using the transformation of Wainscoat \& Cowie (1992), K = 1.282  K$^{\prime}$ $-$
0.282 H.
}
\end{deluxetable}

\clearpage
\begin{table}
\caption{MLCS Parameters for Type Ia Supernovae$^a$}
\vspace{2 mm}
\begin{tabular}{ccccccc}
\hline
Object & T(B$_{max}$) & $\Delta$ & A$_V$ & m$-$M & D (Mpc) & V$_{CMB}$ (km s$^{-1}$) \\
\hline
SN 1999dk  & 414.30 (0.40) & $-$0.38 (0.08) & 0.20 (0.16) & 34.45 (0.18) &  77.6$^{+6.7}_{-6.2}$ & 4218 \\
SN 1999gp  & 551.1 (0.2)   & $-$0.45 (0.10) & 0.30 (0.19) & 35.67 (0.19) & 136.1$^{+9.8}_{-9.0}$  &  7770 \\
SN 2000bk  & 648.5 (1.0)  &    +0.43 (0.15) & 0.28 (0.20) & 35.53 (0.21) & 127.6$^{+13.0}_{-11.7}$ & 7995 \\
SN 2000ce  & 666.3 (1.0)  &  $-$0.26 (0.17) & 1.67 (0.20) & 34.69 (0.22) & 86.7$^{+9.2}_{-8.4}$ & 5088 \\
\hline
\end{tabular}
\begin{tabular}{l}
$^a$B-band maximum is given as Julian Date {\em minus} 2,451,000. \\
$\Delta$ the number of magnitudes that the object is brighter than (negative values) \\
or fainter than (positive values) the fiducial Type Ia SN. \\
A$_V$ is the total V-band extinction (due to our Galaxy {\em and} due to the host galaxy). \\
m$-$M = 5 log $d$ $-$ 5 is the distance modulus in magnitudes, where $d$ is in pc.\\
D is the distance in megaparsecs. \\
The radial velocities (V$_{CMB}$) have been corrected for the motion of the Galaxy toward \\
the Local Group and the motion of the Local Group with respect to the Cosmic Microwave \\
Background Radiation (Kogut et al. 1993). 
\end{tabular}
\end{table}

\begin{table*}
\caption{Light Curve Solutions Following Phillips et al. (1999)$^a$}
\vspace{2 mm}
\begin{tabular}{cccccccc}
\hline
\hline
Object & T(B$_{max}$) & $\Delta$ m$_{15}$(B) & B$_{max}$ & V$_{max}$ 
& I$_{max}$ & E(B$-$V)$_{host}$ & m$-$M \\
\hline
SN 1999da & 370.3(0.6) & 1.94(10) & 16.84(06) & 16.11(06) & 15.85(04) & 0.00(05) & \\
SN 1999dk & 414.5(0.2) & 1.00(10) & 15.06(05) & 14.91(05) & 15.22(05) & 0.07(03) & 34.24(09) \\
SN 1999gp & 552.2(0.1) & 1.00(10) & 16.23(05) & 16.11(05) & 16.34(05) & 0.07(03) & 35.40(09) \\
SN 2000bk & 647.9(0.2) & 1.63(10) & 16.98(20) & 16.81(15) & 16.81(15) & 0.10(04) & 35.48(23) \\
SN 2000ce & 665.7(2.3) & 0.99(10) & 17.24(20) & 16.63(16) & 16.12(19) & 0.54(04) & 34.32(14) \\
\hline
\end{tabular}
\begin{tabular}{l}
$^a$The values in parentheses and without a decimal point are uncertainties in \\
hundredths of a magnitude.
\end{tabular}
\end{table*}


\begin{figure*}
\psfig{figure=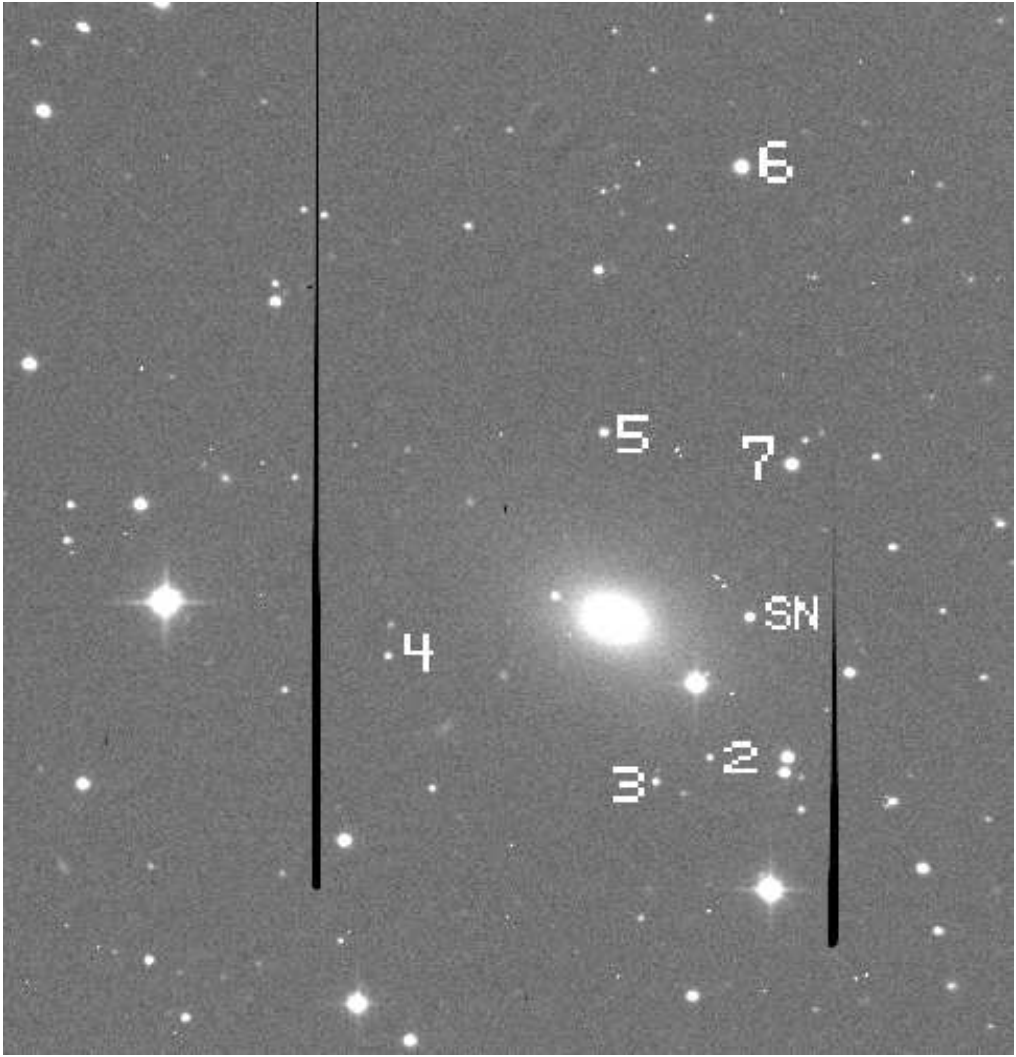,height=14cm}
\caption{A V-band image of NGC 6411 obtained at MRO on 9 July 1999,
with SN 1999da and the stars of the photometric sequence indicated.
The field is 11 arcmin on a side.  North is up, east to the left.}
\end{figure*}

\begin{figure*}
\psfig{figure=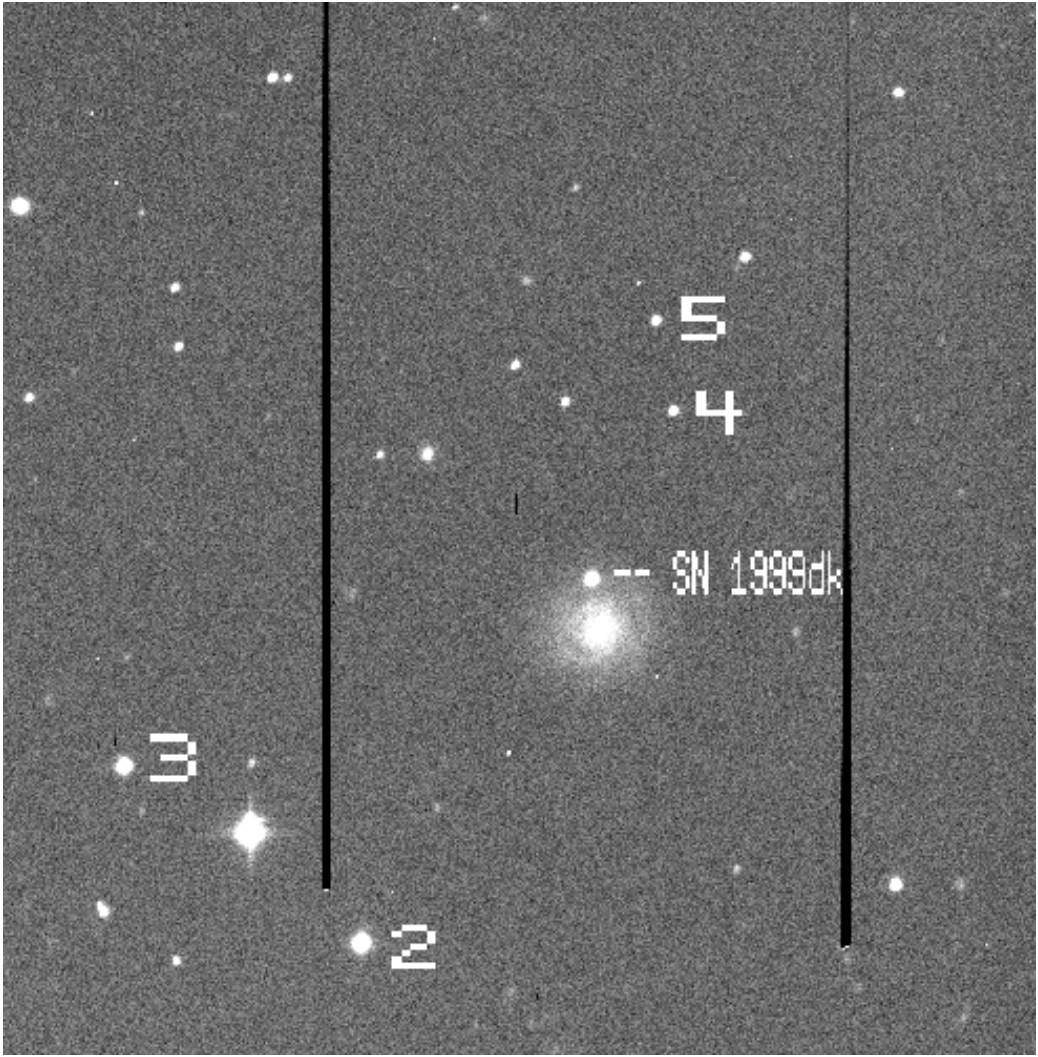,height=14cm}
\caption{A V-band image of UGC 1087 obtained at MRO on 23 August 1999,
with SN 1999dk and the stars of the photometric sequence indicated.
The field is 11 arcmin on a side.  North is up, east to the left.}
\end{figure*}

\begin{figure*}
\psfig{figure=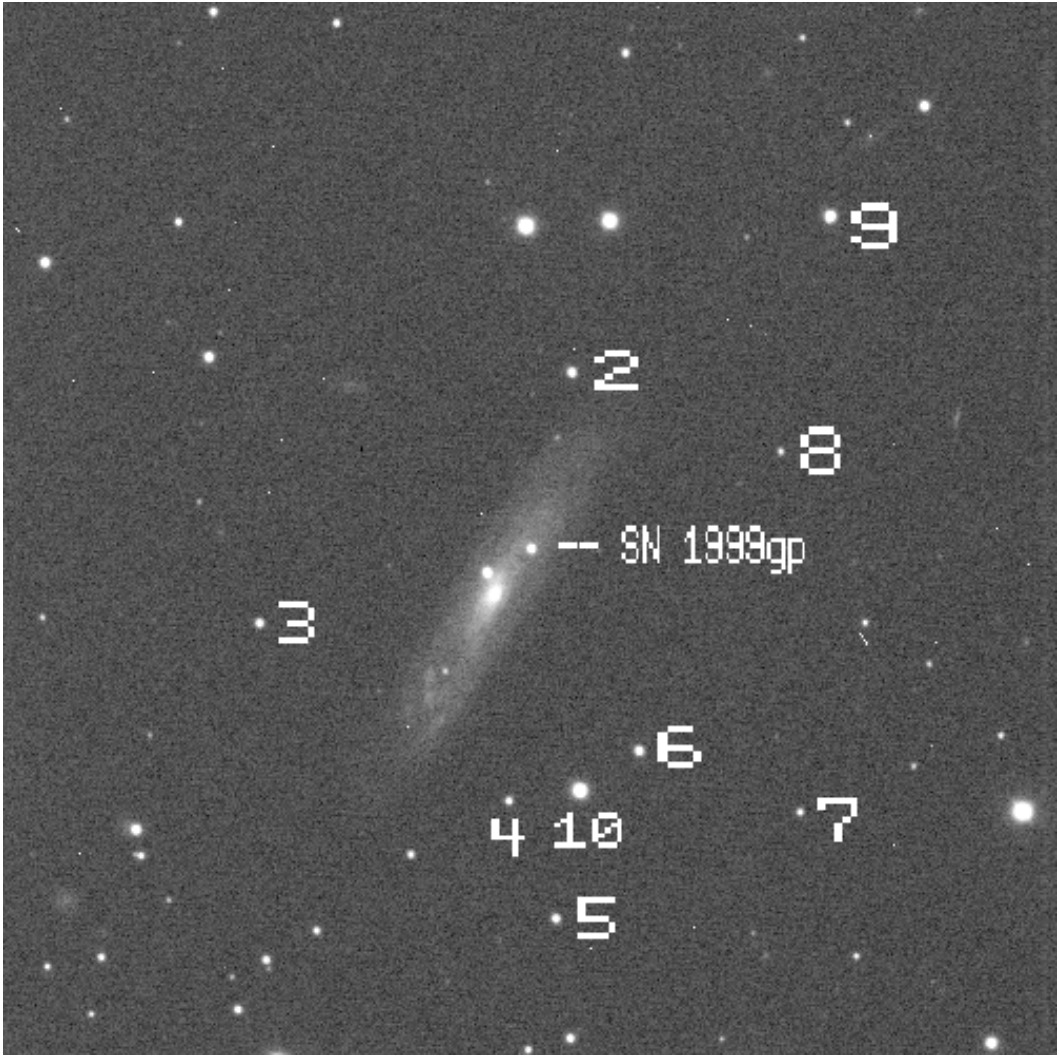,height=14cm}
\caption{A V-band image of UGC 1993 taken on 30 December 1999,
with SN 1999gp and the stars of the photometric sequence indicated.
The field is 4.8 arcmin on a side. The image is rotated 10 degrees
counterclockwise from north up, east to the left.}
\end{figure*}

\begin{figure*}
\psfig{figure=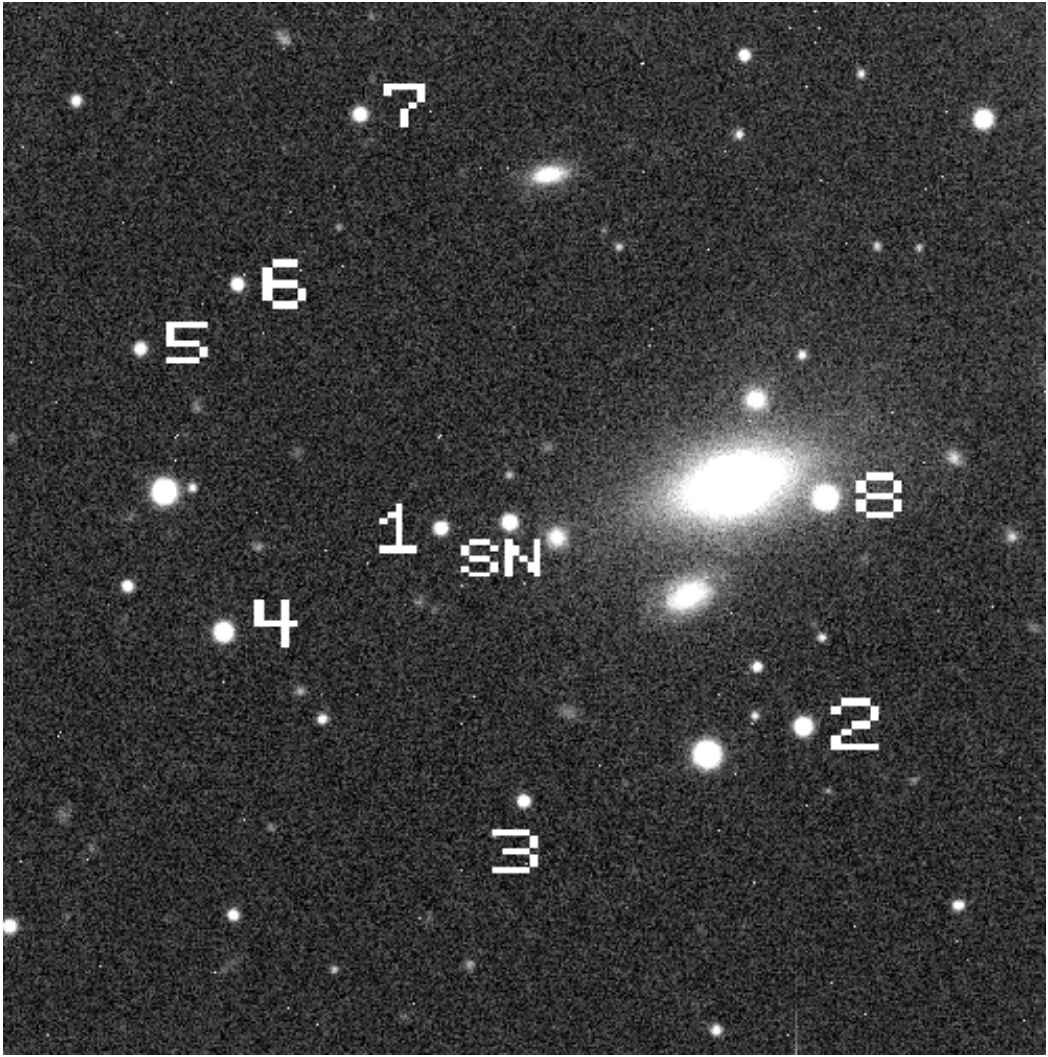,height=14cm}
\caption{A V-band image of NGC 4520 taken on 24 April 2000,
with SN 2000bk and the stars of the photometric sequence indicated.
The field is 4.8 arcmin on a side.  North is up, east to the left.} 
\end{figure*}

\begin{figure*}
\psfig{figure=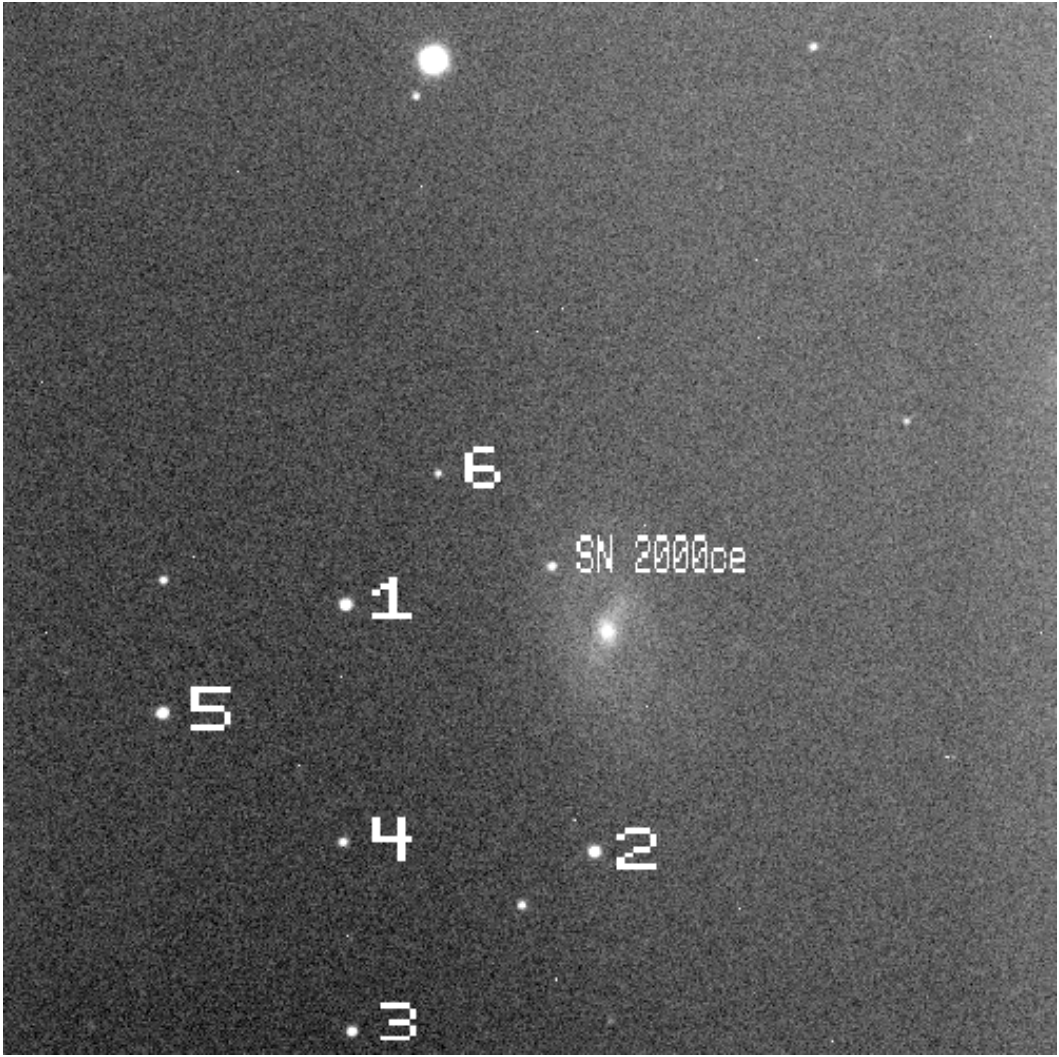,height=14cm}
\caption{A V-band image of UGC 4195 taken on 22 May 2000,
with SN 2000ce and the stars of the photometric sequence indicated.
The field is 4.8 arcmin on a side.  North is up, east to the left.} 
\end{figure*}

\begin{figure*}
\epsfysize=14cm
\centerline{\epsffile{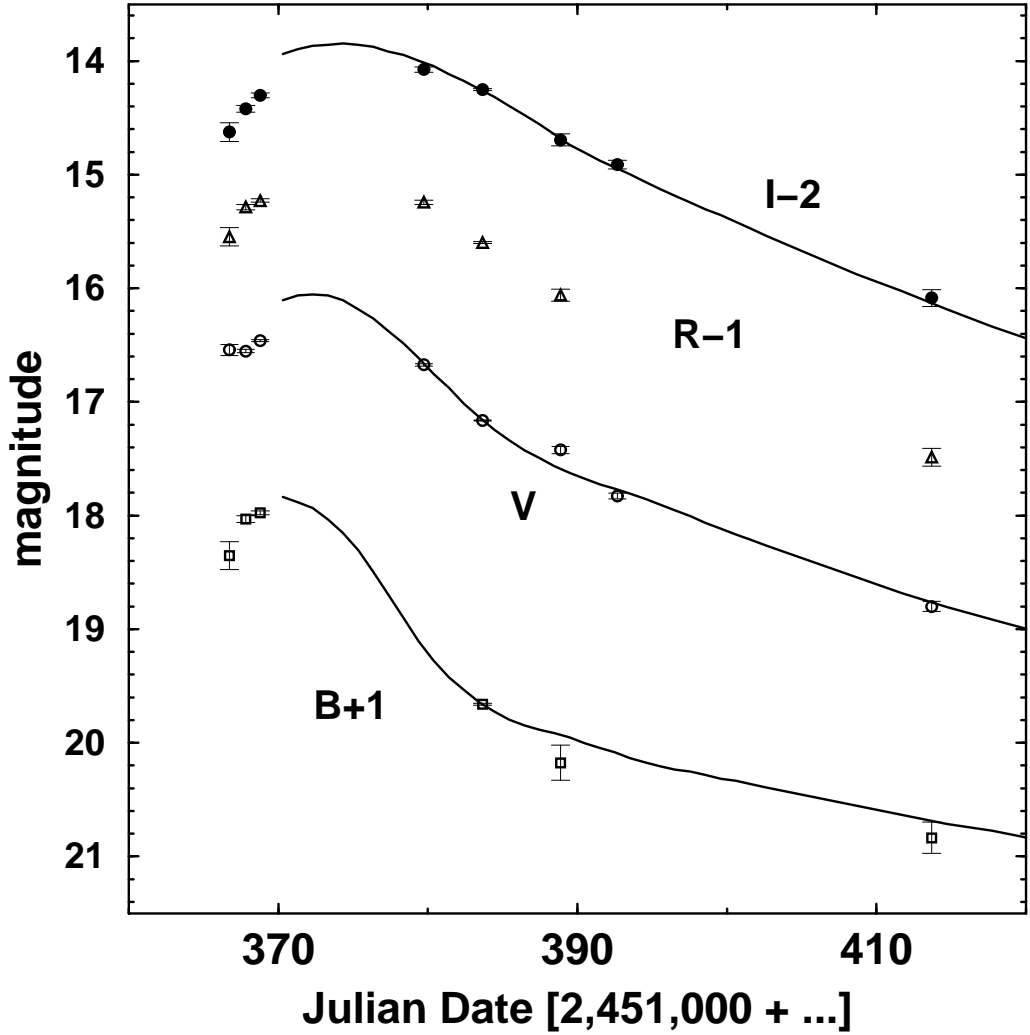}}
\caption{BVRI photometry of SN 1999da.  The B, R, and I
data have been offset vertically by +1, $-$1, and $-$2,
magnitudes, respectively.  
The BVI fits  are based on templates
from the $\Delta$m$_{15}$(B) method of Phillips et al. (1999).}
\end{figure*}

\begin{figure*}
\epsfysize=14cm
\centerline{\epsffile{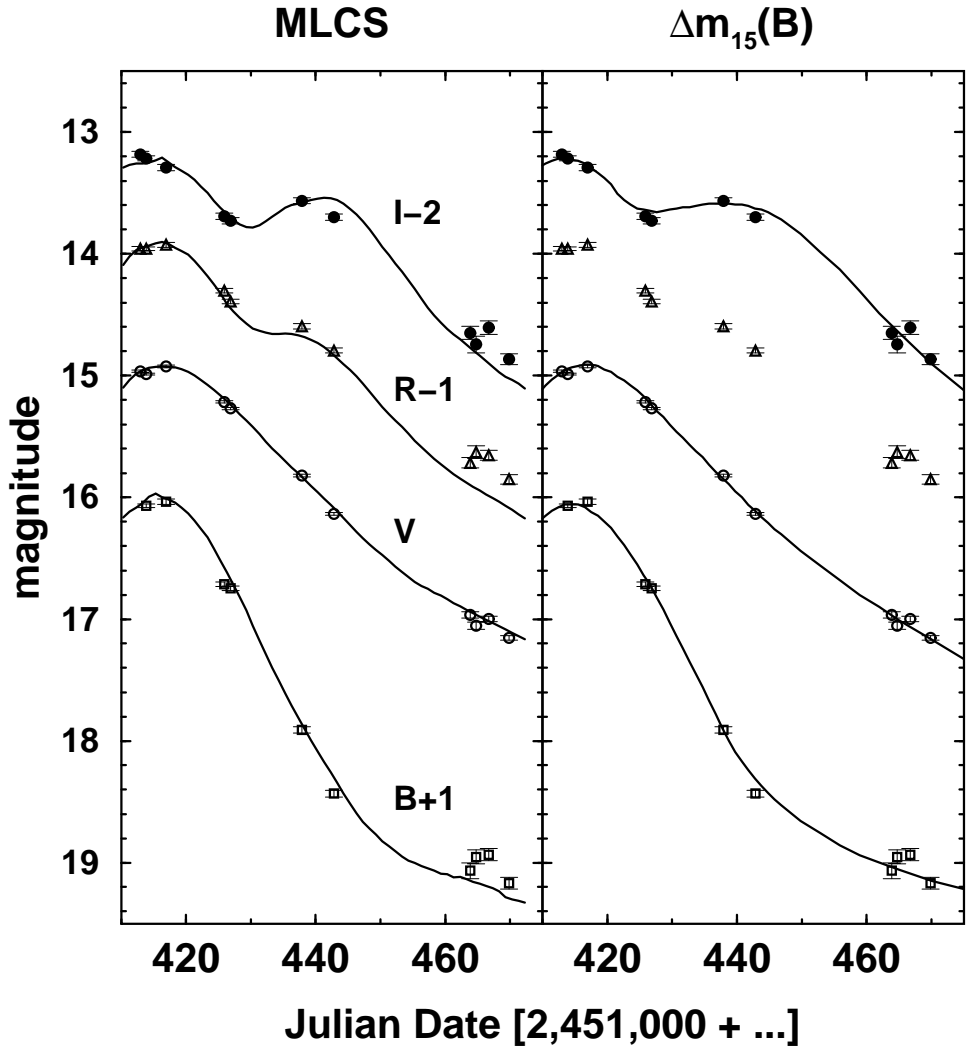}}
\caption{BVRI photometry of SN 1999dk.  The B, R, and I
data have been offset vertically by +1, $-$1, and $-$2,
magnitudes, respectively.  The solid lines in the left
hand panel are based on MLCS v/2.0 
empirical fits with $\Delta$ = $-$0.38 mag.
The BVI fits in the right hand panel are based on templates
from the $\Delta$m$_{15}$(B) method of Phillips et al. (1999).}
\end{figure*}

\clearpage

\begin{figure*}
\epsfysize=14cm
\centerline{\epsffile{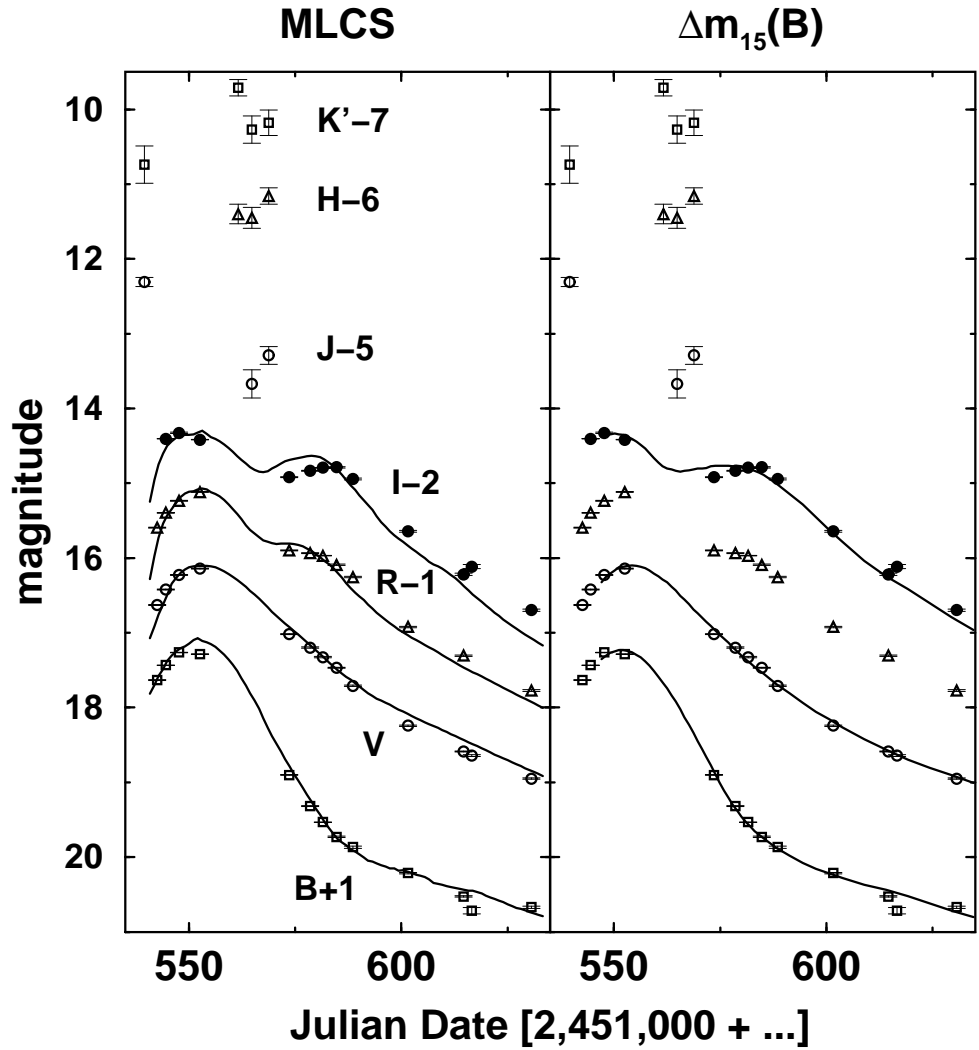}}
\caption{BVRI photometry of SN 1999gp.  The B, R, I, J, H,
and K$^{\prime}$ data have been offset vertically by +1, $-$1, $-$2,
$-$5, $-$6, and $-$7 magnitudes, respectively.  MLCS v/2.0 templates with
$\Delta = -0.45$ are shown in the left hand panel.
The BVI fits in the right hand panel are based on templates
from the $\Delta$m$_{15}$(B) method of Phillips et al. (1999).}
\end{figure*}

\begin{figure*}
\epsfysize=14cm
\centerline{\epsffile{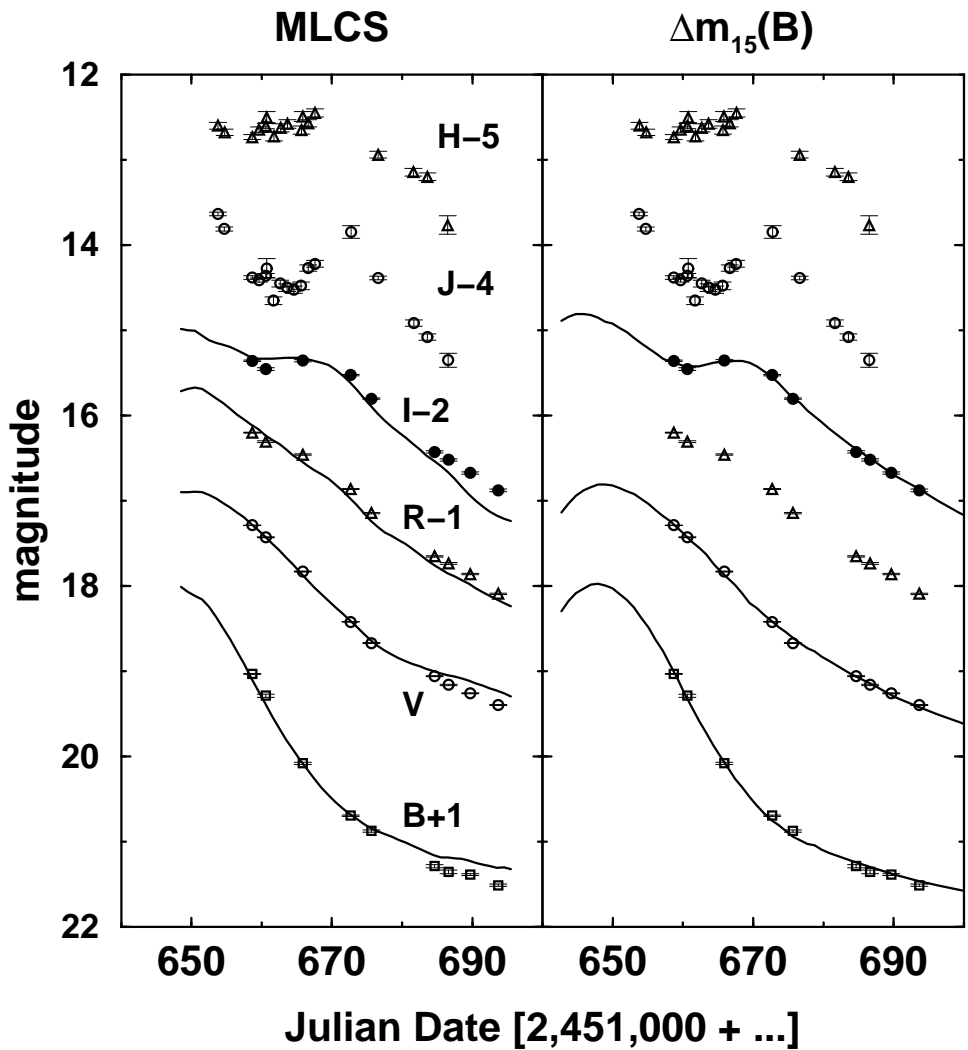}}
\caption{BVRI and near infrared photometry of SN 2000bk, with MLCS v/2.0 fits
using $\Delta$ = +0.43 shown in the left hand panel.
The BVI fits in the right hand panel are based on templates
from the $\Delta$m$_{15}$(B) method of Phillips et al. (1999).
The B, R, I, J, and H data have
been offset vertically by +1, $-$1, $-$2, $-$4, and $-$5 magnitudes,
respectively. Most of the J-band photometry was actually obtained 
with a J$_s$ filter.}
\end{figure*}

\begin{figure*}
\epsfysize=14cm
\centerline{\epsffile{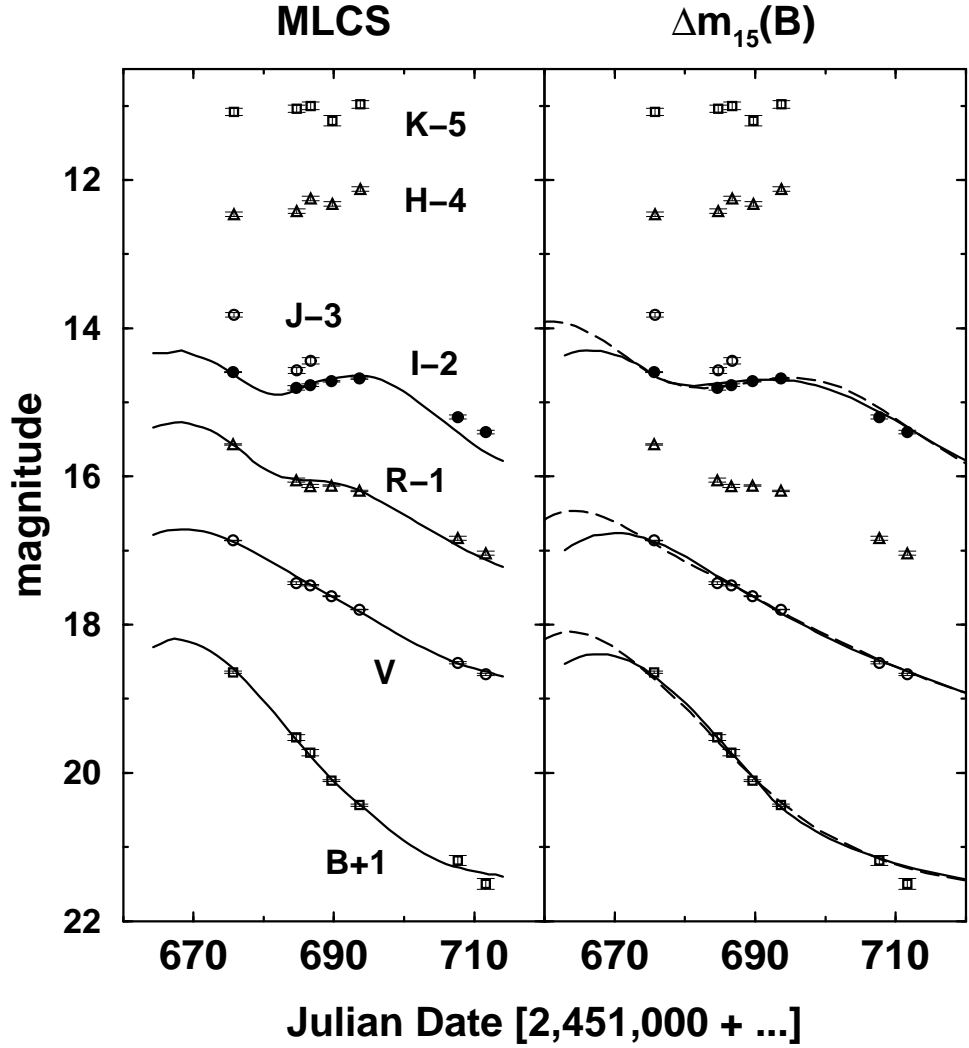}}
\caption{BVRI photometry of SN 2000ce, with MLCS v/2.0 fits using $\Delta 
= -0.26$ shown in the left hand panel. Two templates from the
$\Delta$m$_{15}$(B) method of Phillips et al. (1999) are shown
in the right hand panel and fit the data equally well.  This
implies a greater than normal uncertainty in the time of maximum
light and the magnitudes at maximum. The B, R, I, J, H, and K data
have been offset vertically by +1, $-$1, $-$2, $-$3, $-$4, and $-$5
magnitudes, respectively.}
\end{figure*}

\begin{figure*}
\epsfysize=14cm
\centerline{\epsffile{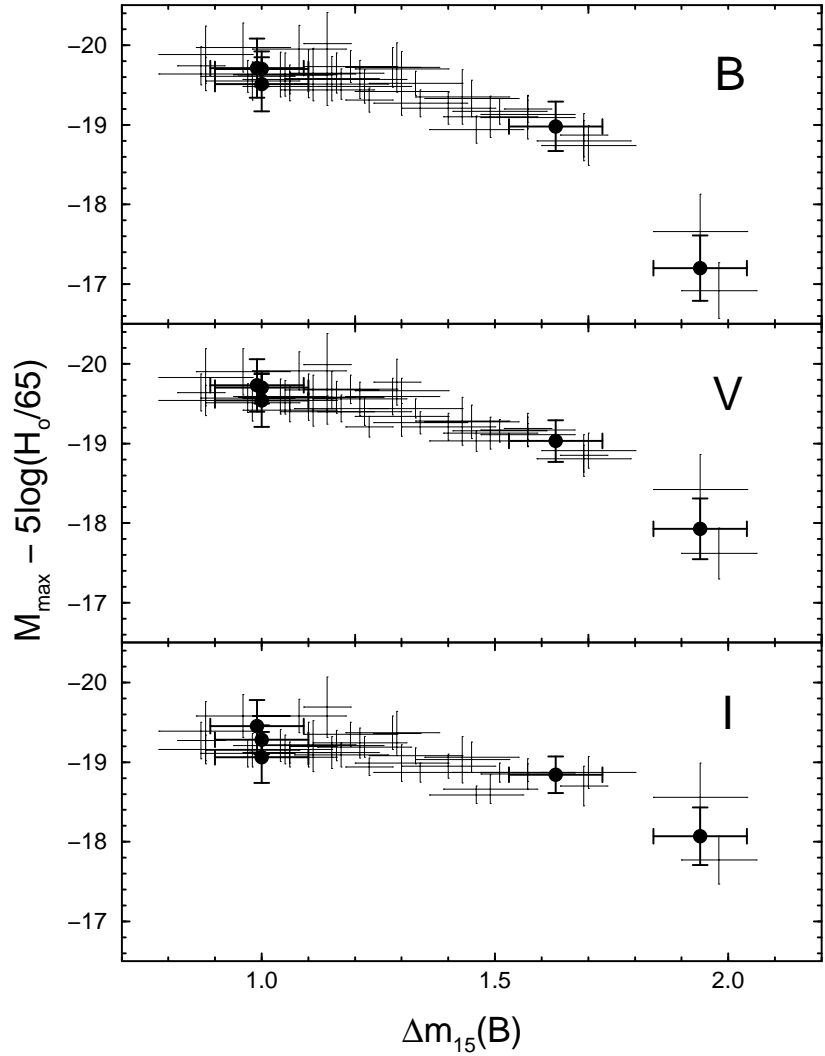}}
\caption{Absolute magnitudes at maximum for the 5 SNe from this
paper (filled symbols), plus 41 SNe from Phillips et al.
(1999), and SN 1998de (Modjaz et al. 2001). All of these objects
have redshift $z \gtrsim$ 0.01, so the effect of peculiar motions
should not significantly contribute to the scatter.}
\end{figure*}

\begin{figure*}
\psfig{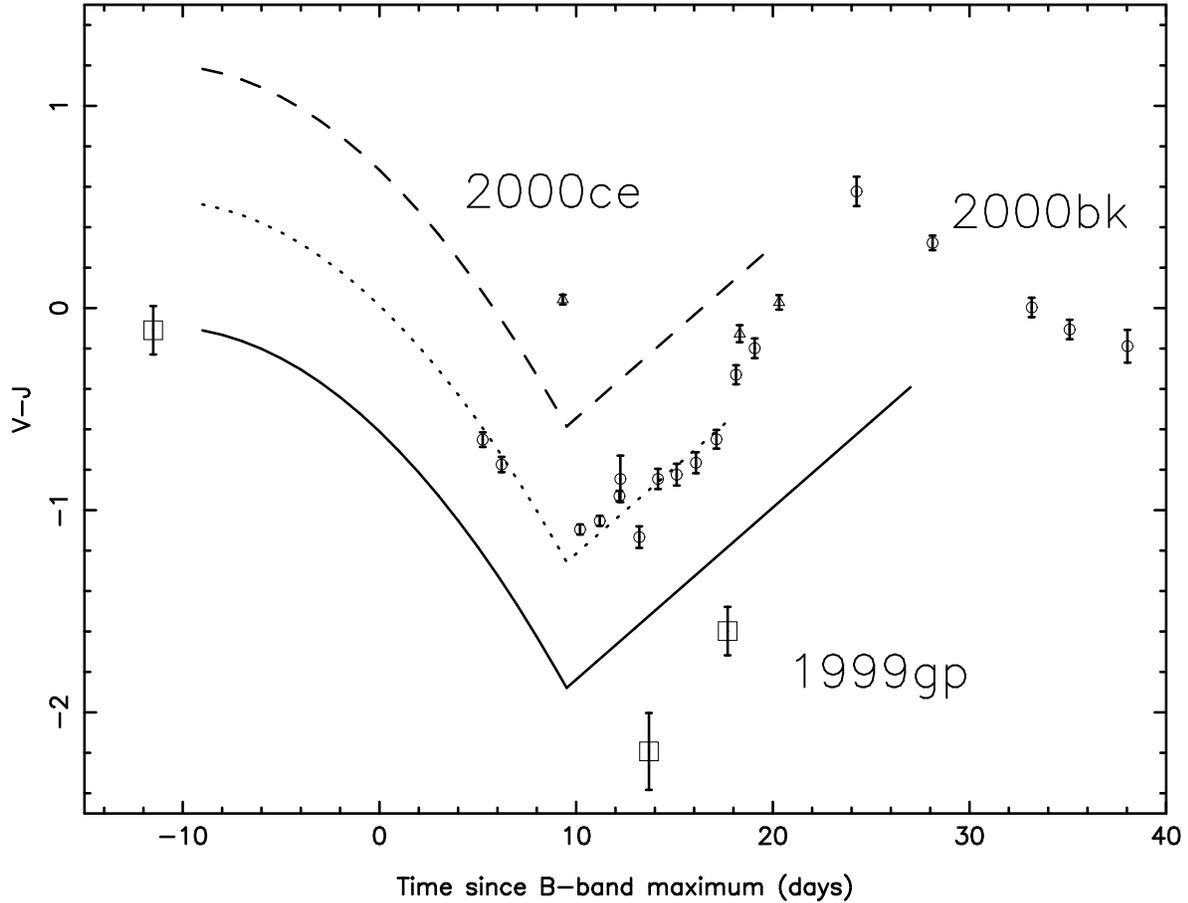}
\caption{Observed V$-$J colors of SNe 1999gp (squares),
2000bk (open circles), and 2000ce (triangles). The solid line is the V$-$J
unreddened locus from Table 9 of Krisciunas 
et al. (2000). The dotted line is the unreddened locus offset
by 0.62 mag.  Clearly, a simple upward translation of the unreddened locus
does not fit the shape of the V$-$J color curve for SN 2000bk beyond $t$ = 18
days. The dashed line is the unreddened locus offset by 1.29 mag;
it only fits the SN 2000ce points well if the points are shifted 3.0 days
to the left.}
\end{figure*}

\begin{figure*}
\psfig{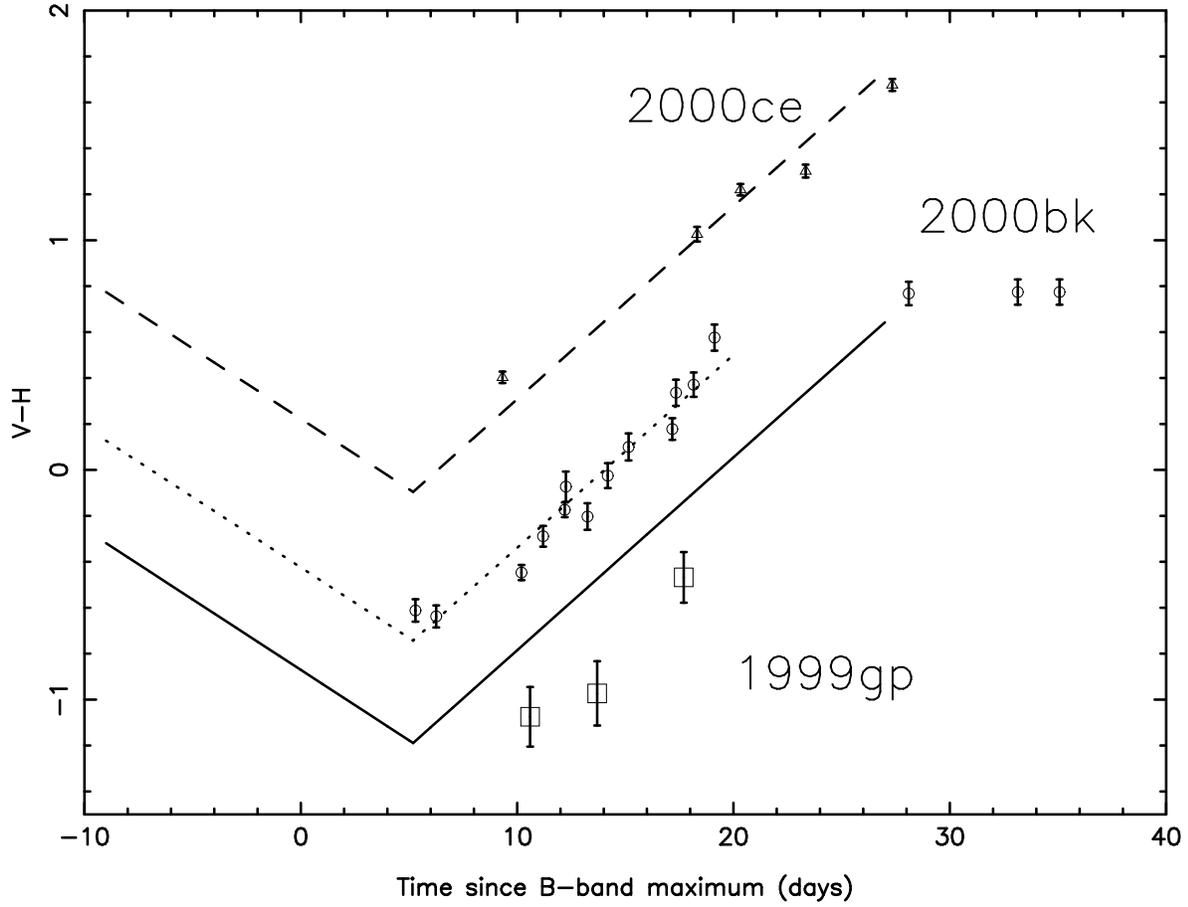}
\caption{Observed V$-$H colors of SNe 1999gp,
2000bk, and 2000ce. The solid line is the V$-$H 
unreddened locus from Table 9 of Krisciunas 
et al. (2000). The dotted line is the unreddened locus offset
by 0.45 mag. The dashed line is the unreddened locus offset by 1.09 mag.}
\end{figure*}

\begin{figure*}
\psfig{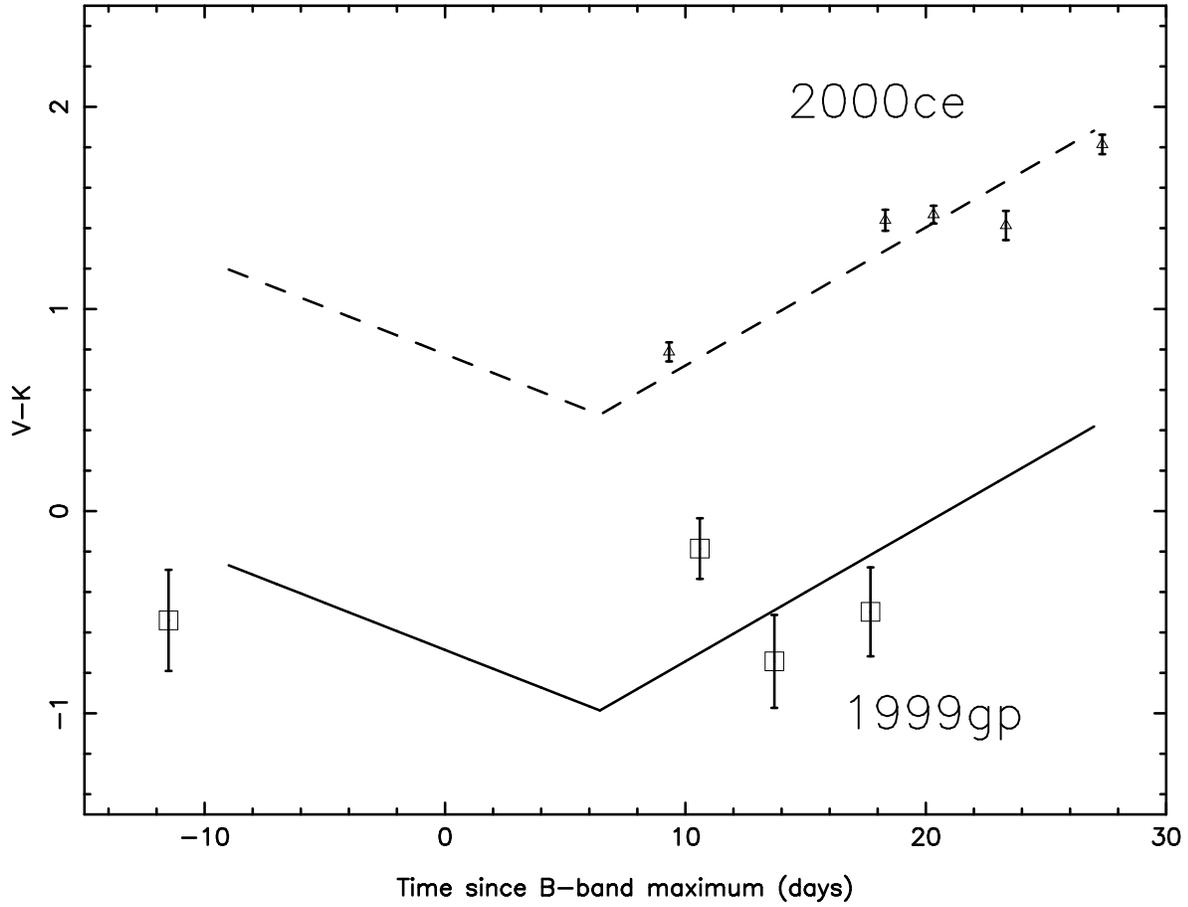}
\caption{Observed V$-$K colors of SNe 1999gp and 2000ce.
The solid line is the V$-$K unreddened locus from Table 9
of Krisciunas et al. (2000).  The
dashed line is the unreddened locus offset by 1.46 mag.
The leftmost point for SN 1999gp is actually V$-$K$^\prime$.}
\end{figure*}

\begin{figure*}
\psfig{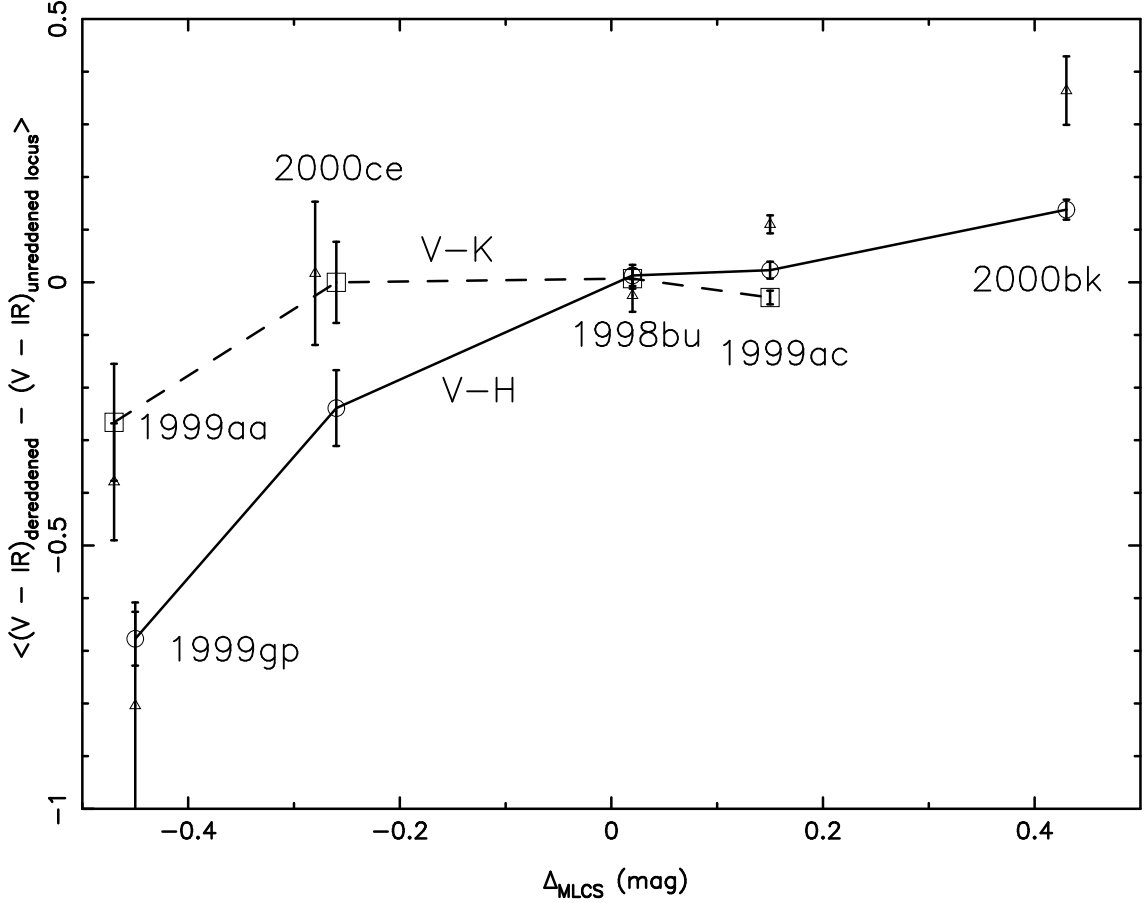}
\caption{We have dereddened the V$-$J (triangles), 
V$-$H (circles) and V$-$K (squares)
colors for several supernovae and determined the mean residuals
of those data with respect to unreddened loci derived from data
used for Paper I.  Except for SN 1998bu we have restricted the
data to 0 $\leq t \leq$ 27 days after the time of B-band maximum. 
We have horizontally offset the J-band point for SN 2000ce for plotting
purposes.  The solid line and dashed line are naive 
connect-the-dots ``fits'' to the V$-$H and V$-$K data, respectively.
This figure shows that the mean residual is
near zero for V$-$K over a range of luminosity-at-maximum.
Overluminous, slowly declining supernovae like
1999aa and 1999gp are intrinsically bluer in V$-$ near IR color 
indices than more rapidly declining objects. More rapidly declining
objects like SN 2000bk have redder colors than the mid-range decliners.}
\end{figure*}

\begin{figure*}
\psfig{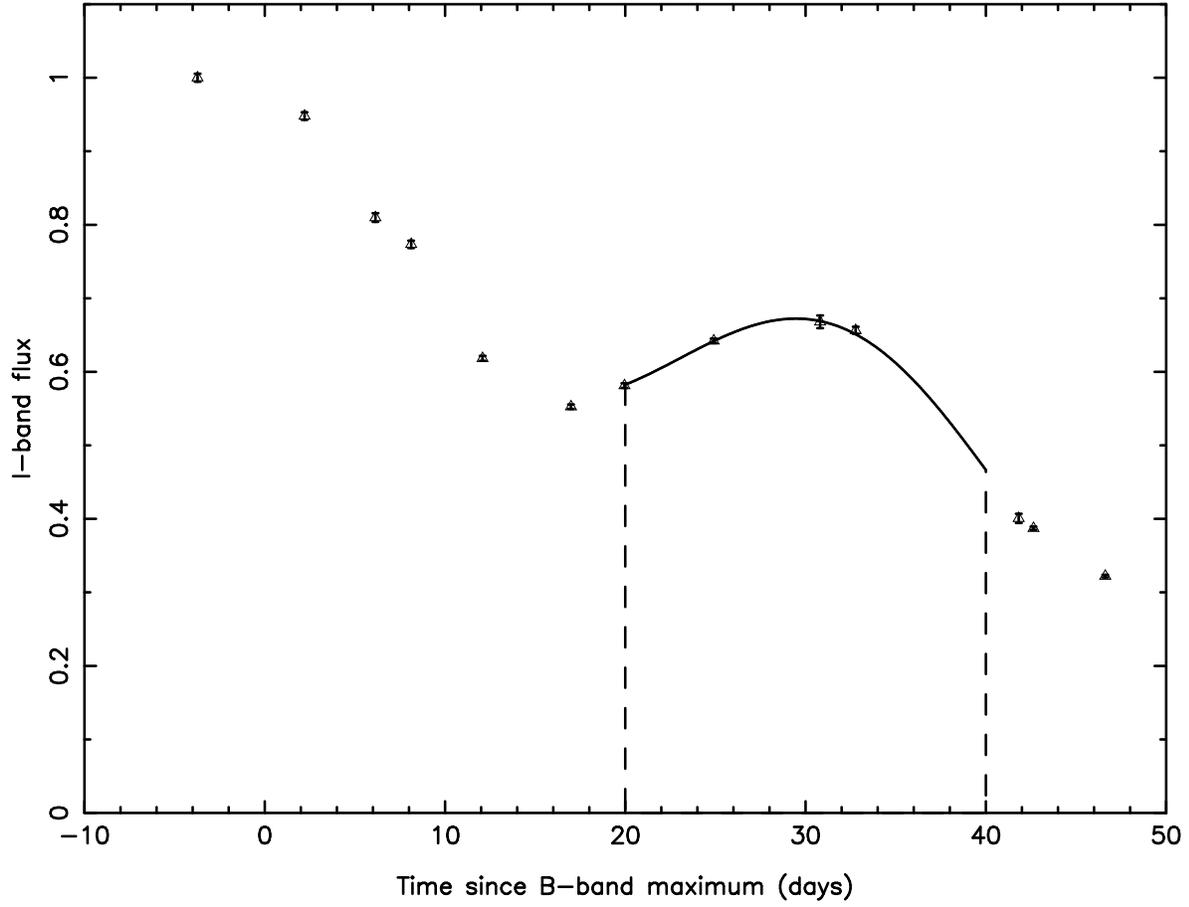}
\caption{I-band light curve of SN 1999aa, with the magnitudes
converted to flux units and normalized to the maximum brightness.
An integration of a polynomial fit to the data allows us 
to determine the mean flux from 20 to 40 days after the 
time of B-band maximum, $\langle$ I $\rangle _{20-40}$.}
\end{figure*}

\begin{figure*}
\psfig{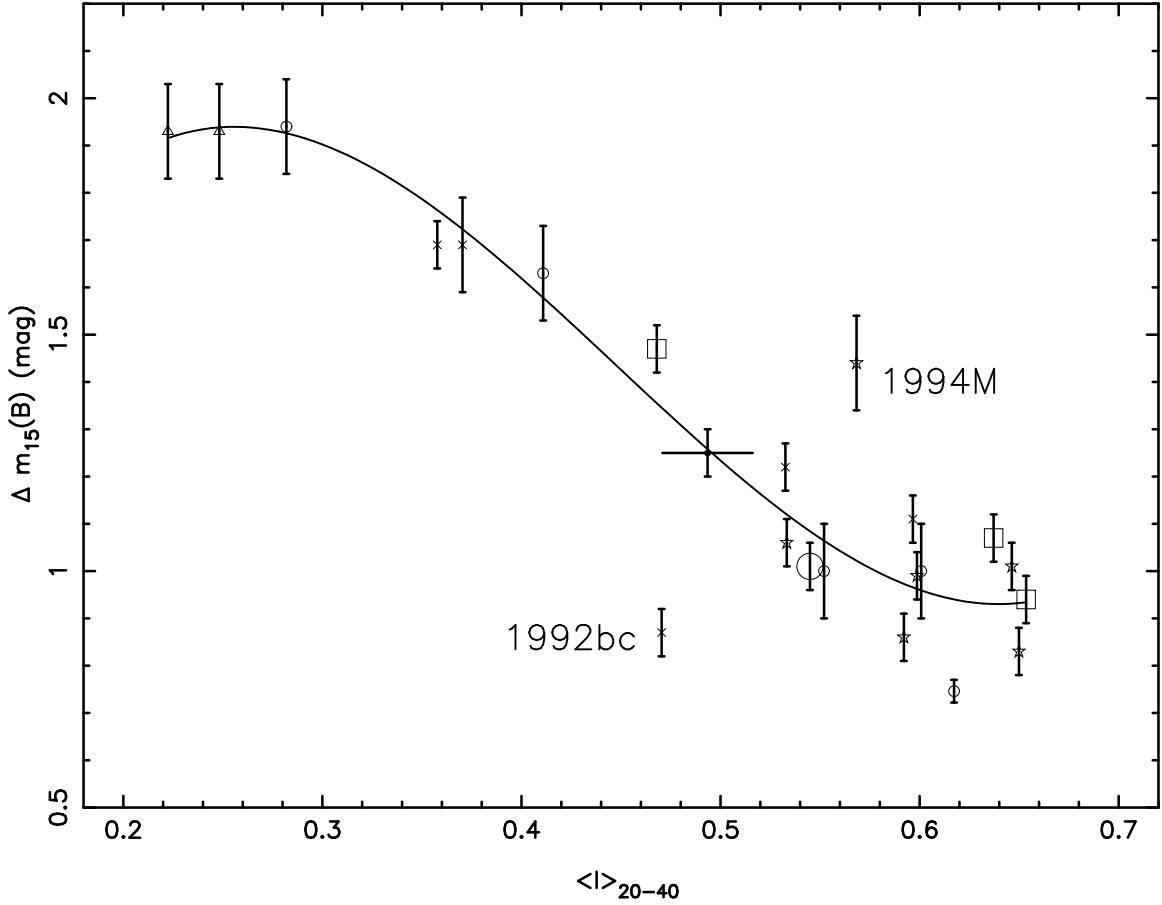}
\caption{The decline rate parameter $\Delta$m$_{15}$(B) vs.
the mean I-band flux 20 to 40 days after
the time of B-band maximum.  Symbols: squares $-$ three SNe
from the RPK training set; triangles $-$ SNe 1992K and 1991bg; large
open circle $-$ SN 1998bu; X's $-$ five SNe from the
Cal\'{a}n/Tololo survey; five pointed stars $-$
six SNe from Riess et al. (1999); small open circles $-$ 
four SNe from this paper, plus SN 1999aa (Paper I); the point
with the horizontal error bar represents the mean of the values
derived from the SN 1996X data sets of Riess et al. (1999) and
R. Covarrubias et al. (unpublished). The two
objects which are found furthest from the third order regression line
are labeled.}
\end{figure*}

\begin{figure*}
\epsfysize=14cm
\centerline{\epsffile{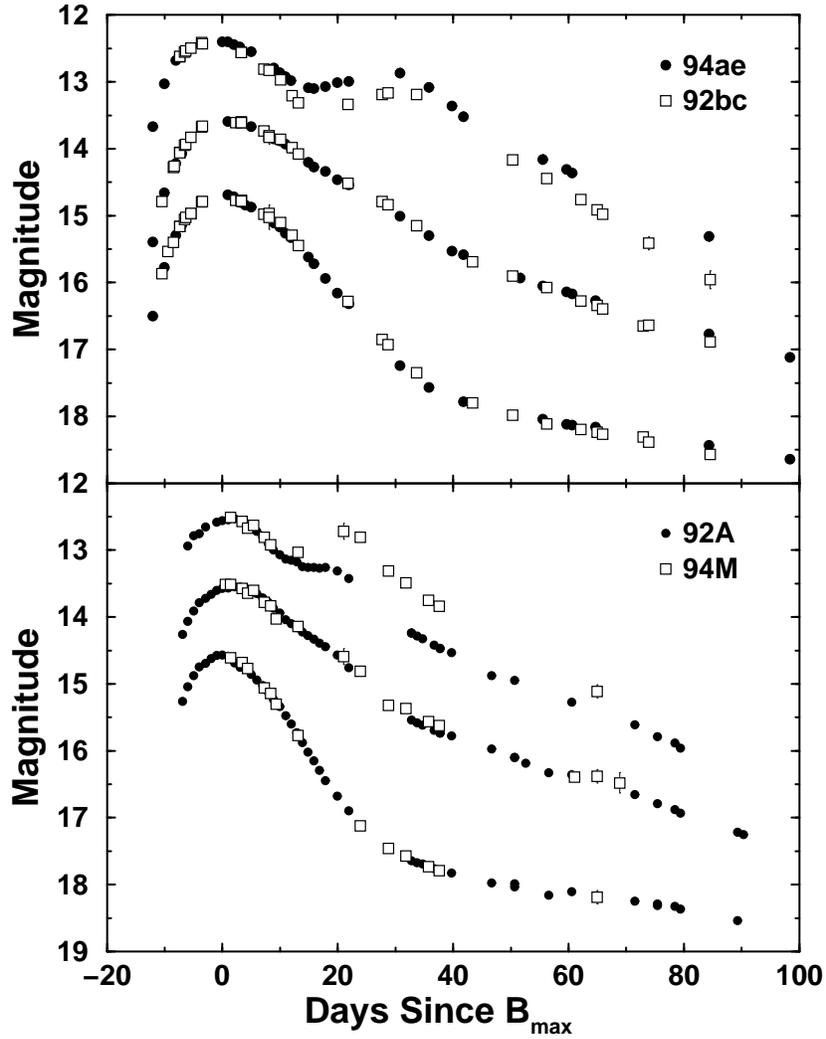}}
\caption{In each panel we show, from the top down, the I-, V-, and
B-band light curves of Type Ia supernovae, with the photometry 
adjusted to make the maxima coincide.  These two pair of supernovae
illustrate that objects can have identical decline rates in B and
V, yet have greatly different I-band secondary maxima.}
\end{figure*}

\end{document}